\documentclass{article}
\usepackage[utf8]{inputenc}

\usepackage{graphicx}        

\usepackage{amsmath, amssymb}
\usepackage{hyperref}
\usepackage{booktabs}
\usepackage[authoryear]{natbib}
\usepackage[english]{babel}
\usepackage{authblk}
\usepackage{geometry}
\usepackage[dvipsnames]{xcolor}
\usepackage{enumitem}
\usepackage{rotating}

\hypersetup{
			colorlinks=true,
			linkcolor=black,
            linktoc=page,
			anchorcolor=black,
			citecolor=blue,
			urlcolor=black,
}

\def \bs{\mathbf}
\def\diag{\mbox{diag}}
\def \bqmatrix{\begin{pmatrix}}
\def \eqmatrix{\end{pmatrix}}
\newtheorem{proposition}{Proposition}
\def \green{\textcolor{black}}
\def \red{\textcolor{black}}

\begin{document}
\title{\bf Quantile Mixed Hidden Markov Models for multivariate longitudinal data
}
\author[1]{Luca Merlo}
\affil[1]{Department of Statistics, Sapienza University of Rome}

\author[2]{Lea Petrella}
\affil[2]{MEMOTEF Department, Sapienza University of Rome}

\author[3]{Nikos Tzavidis}
\affil[3]{Department of Social Statistics and Demography, Southampton Statistical Sciences Research Institute, University of Southampton}


%
%
\maketitle

\abstract{
\red{The identification of factors associated with mental and behavioral disorders in early childhood is critical both for psychopathology research and the support of primary health care practices. 
 Motivated by the Millennium Cohort Study, 
 in this paper we study the effect of a comprehensive set of covariates on children's emotional and behavioural trajectories 
 in England. To this end, 
 we develop} a Quantile Mixed Hidden Markov Model for joint estimation of multiple quantiles in a linear regression setting for multivariate longitudinal data. The \green{novelty of the proposed} approach is based on the Multivariate Asymmetric Laplace distribution which allows to jointly estimate the quantiles of the univariate conditional distributions of a multivariate response, accounting for possible correlation between the outcomes.
 \green{Sources of unobserved heterogeneity} and serial \red{dependency} due to repeated measures  
are modeled through the introduction of individual-specific, time-constant random coefficients and time-varying parameters evolving over time with a Markovian structure, respectively. The inferential approach is carried out through the construction of a suitable Expectation-Maximization algorithm without parametric assumptions on the random effects distribution. 
 \\}  



\section{Introduction}\label{sec:intro}
\red{The occurrence of stressful life events, family environment and poverty are key contributing risk factors to children's emotional and behavioural disorders (\cite{bradley2002socioeconomic, goodman2003using, flouri2010area, goodnight2012quasi, platt2016stressful}). Mental health problems at an early age can create considerable distress for the child and the family and can have a significant impact on the child’s social, emotional and psychological development. Therefore, accurate identification of mental disorders is \red{important} for psychologists and clinicians in order to reduce this disruption, avoiding that psychological problems will persist into adulthood, and to better understand the problem for timely treatment recommendations (\cite{becker2004evaluation, mathai2004comparing, van2008parent}). This is particularly for children \green{facing major problems} where, as risk factors accumulate, emotional and behavioural problems tend to increase considerably (\cite{trentacosta2008relations}).}
 \red{In this context,} one of the most widely and internationally used measure of child mental health is provided by the Strengths and Difficulties Questionnaire (SDQ) (see \cite{goodman1997strengths} and \cite{goodman2009strengths}). It \red{offers} a balanced coverage of children and young people's behaviours, emotions and relationships and it has been designed to measure children’s emotional and behavioural problems in psychological research.
 \red{SDQ outcomes are assessed by diagnostic measures of child disorders, and collected by parent- and teacher- reported measures of personal, emotional and social development.}
 On one hand, internalizing behaviours are \red{manifested} by inward symptoms such as being withdrawn, fearful or anxious; on the other hand, externalizing behaviours are outward and may be described as aggressive, non-compliant, impulsive or fidgety. The SDQ score is the sum of the main caregiver's responses to a series of items that describe children's internalizing and externalizing problems. This covers five different domains: emotional symptoms, peer problems, conduct problems, hyperactivity, and pro-social behavior. Each domain is measured by five items, for a total of 25 items. For each item, a score equal to 0 is given if the response is not true, 1 if it is somewhat true and 2 if it is certainly true. The internalizing SDQ score is the sum of the scores for responses to the five items in the domains of emotional and peer problems while the externalizing SDQ score is the sum of responses to the five conduct problems and hyperactivity items. Therefore, both SDQ scores range from 0 to 20 
 \red{
 (for further details refer to \href{https://www.sdqinfo.com/}{www.sdqinfo.com}).
\\
 In this paper we focus on emotional and behavioural disorders of children who participated in the Millennium Cohort Study (MCS),
 which is a longitudinal birth cohort study following children born 
in the United Kingdom (UK), providing multiple measures of the cohort members' physical, socio-emotional, cognitive and behavioural development over time. An extensive literature has examined and documented the effect of mother's characteristics, neighbourhood context and family risk factors on children's trajectories SDQ scores collected by the MCS (see, for example, \cite{mcmunn2012maternal, tzavidis2016, flouri2016prosocial, wickham2017effect, alfo2020m}). Thanks to the longitudinal structure of the cohort data, these studies can shed light on the evolution of SDQ scores over time and on how they are affected by risk factors and other family and child characteristics.}
 
\begin{figure}[!h]
\center
\includegraphics[width=1\linewidth, height=6.0cm, keepaspectratio]{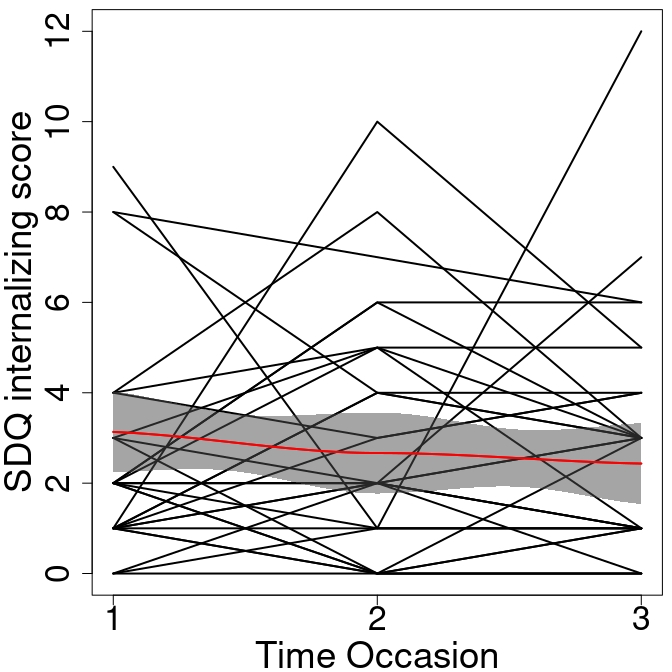}
\includegraphics[width=1\linewidth, height=6.0cm, keepaspectratio]{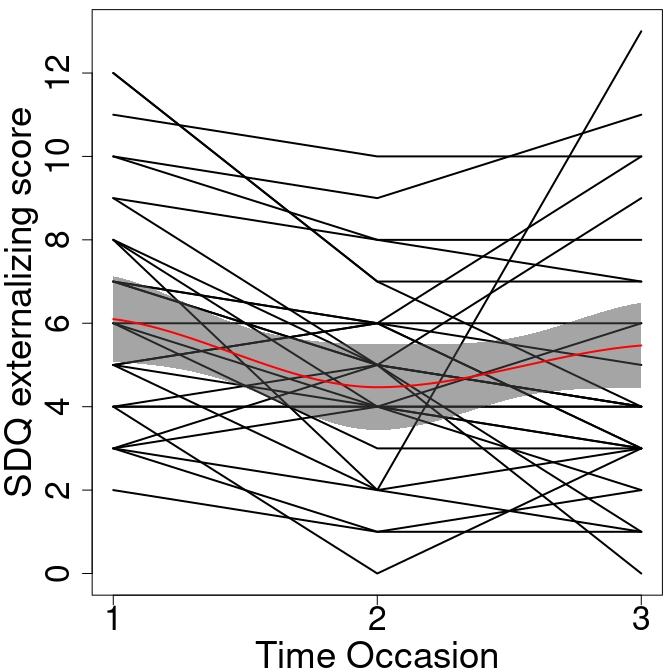}
\caption{Individual trajectories for a random subsample of 30 children for SDQ internalizing (left) and externalizing (right) problems.}\label{fig:ind}
\end{figure}

 \red{The analysis of SDQ data poses crucial challenges for statistical modeling. As shown in Figure \ref{fig:data}, the distributions of SDQ scores are non-negative, positively skewed and exhibit atypical values. 
 Typically, linear random effect models (\cite{laird1982random, lindsey1999models, diggle2002analysis, goldstein2011multilevel}) have been implemented with much of the focus being univariate and centered on the conditional mean of the distribution given a set of covariates. Nevertheless, modeling the conditional
mean may not offer the best summary as linear models perform badly with non-Gaussian data. 
  Moreover, it is possible that the effect of certain risk factors on the SDQ scores is not the same across \red{the SDQ distribution}. Indeed, there is empirical evidence to suggest that socio-economic and parental factors have a more pronounced \red{effects} at the top end where children display a high, perhaps abnormal, level of adjustment problems than at the bottom end of the distribution (see e.g. \cite{kiernan2008economic, tzavidis2016}). It is therefore important to identify the predictors of children's disorders, not only at the average but \red{more importantly, in the upper part of the SDQ distribution as this relates} to high-risk youths having elevated levels of mental health and conduct problems. 
\begin{figure}[!h]
\center
\includegraphics[width=1\linewidth, height=6.7cm, keepaspectratio]{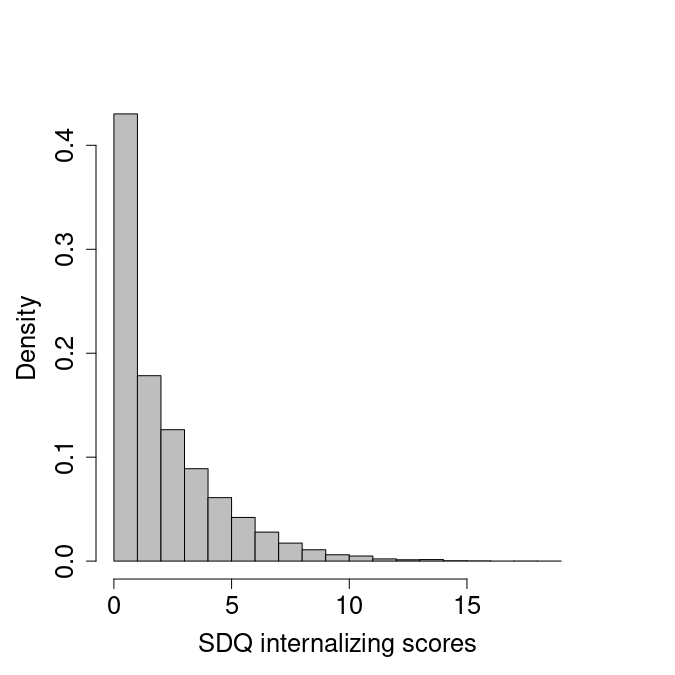}
\includegraphics[width=1\linewidth, height=6.7cm, keepaspectratio]{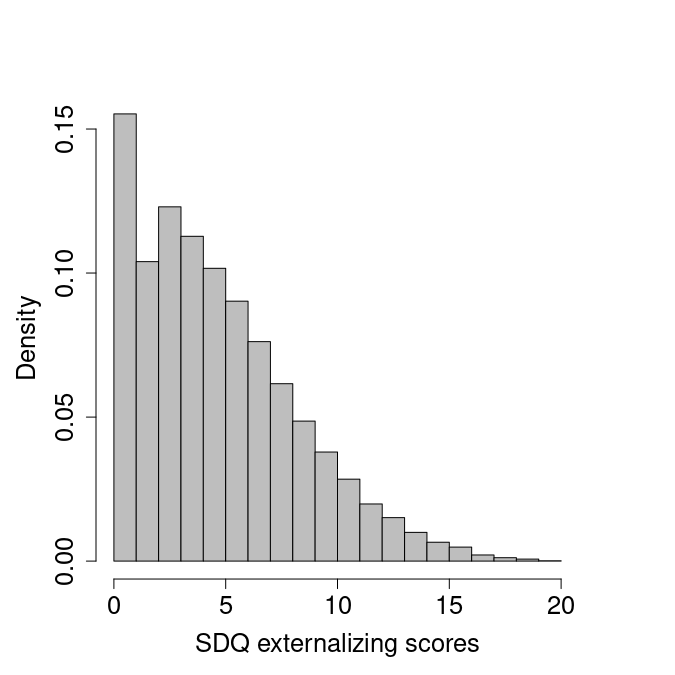}
\caption{Histogram for SDQ internalizing (left) and externalizing (right) problems scores.}\label{fig:data}
\end{figure} 
\\
  In this context, a quantile regression approach can be more appropriate \red{for getting} a complete picture of the entire conditional distribution of children's difficulties than ordinary regression \red{around the mean}.} Quantile regression provides a way to model the conditional quantiles of a response variable with respect to a set of covariates and may reveal how their effect varies at different parts of the response distribution. Quantile regression methods have become widely used in literature because they are suitable in those situations where skewness, fat-tails, outliers, truncation, censoring and heteroscedasticity arise, and they have been implemented in a wide range of different fields, both in a frequentist paradigm and in a Bayesian setting. For a detailed review and list of references see \cite{koenker2005quantile} and \cite{koenker2017handbook}.
\\
\red{Owing to the longitudinal structure of the MCS cohort data \red{(measurements recorded on the same individuals)}, the potential association between dependent observations should be taken into account in order to provide correct inference. Within the quantile regression literature, random effects models have been \red{presented} to accommodate time-constant, within-subject correlation 
and between subject heterogeneity: \cite{liu2009mixed} and \cite{geraci2014linear} proposed to add time-constant individual-specific random coefficients in the regression model to capture unobserved heterogeneity. However, when analyzing internalizing and externalizing problems, it is reasonable to expect that individual temporal trajectories of SDQ scores vary from child to child. Figure \ref{fig:ind} shows the individual trajectories of SDQ scores for a random subset of children, 
where the overall trend, as estimated by a local polynomial regression, is shown in red along with the 95\% confidence bands (highlighted in grey). While the general trend is relatively constant over time, individual trajectories show rapid changes, especially ``U''-shaped curves for externalizing score measurements. For example, trajectories for children exposed to ‘high risk’ circumstances may become less salient with age, whereas for children at ‘low risk’, they could become more salient as they make the transition into adolescence. This possibly translates to decreasing and increasing intercepts in the dynamics of SDQ levels that require specific modeling tools. In this case, time-constant random effects can lead to biased estimates (\cite{bartolucci2009multivariate, farcomeni2012quantile}) meanwhile, their temporal evolution can be better captured by the introduction of time-dependent intercepts.}

 \red{To account for serial heterogeneity, \cite{farcomeni2012quantile} suggested the use of Hidden Markov Models (HMM). In such a context, a latent homogeneous Markov chain is defined in order to capture the temporal evolution of unobserved heterogeneity and 
state-dependent parameters are introduced to account for response variability due to time-varying omitted covariates. 
 \\
 \red{To handle longitudinal data where both time-constant and time-varying sources of unobserved heterogeneity are present,} it is possible to consider the well-known Mixed Hidden Markov Models (MHMM) (see \cite{altman2007mixed}). The MHMM is obtained by combining the features of HMMs and mixed effects models, encompasses linear mixed models and HMMs as it accommodates time-constant and time-varying sources of variability jointly. In the application of quantile regression to longitudinal data, \cite{marino2018mixed} introduced a Mixed Hidden Markov quantile regression model for longitudinal continuous responses, 
extending the linear quantile mixed model of \cite{geraci2014linear} and the linear quantile HMM of \cite{farcomeni2012quantile}. These proposals are, however, designed for univariate dependent variables and consequently, they do not account for the dependence structure between multiple outcomes of interest measured longitudinally.
\\
Another important aspect when analyzing SDQ scores concerns the dependence structure between mental and behavioural problems. In the literature, \cite{lilienfeld2003comorbidity, liu2004childhood, cicchetti2014developmental} and \cite{alfo2020m} have given empirical proof of the existence of correlation between SDQ scores, which places children at a greater risk of developing internalizing-externalizing comorbidity. In this case, univariate approaches completely ignore the correlation structure between children outcomes; by contrast, a \green{multivariate} analysis would be able to identify the underlying \green{drivers} that affect changes in each of the distribution of the SDQ scores and, at the same time, give valuable insight about the correlation among children disorders.}
\\
When multivariate response variables are concerned, however, the univariate quantile regression method does not straightforwardly extend to higher dimensions 
 since there is no “natural” ordering in a $p$-dimensional space, for $p>1$.
As a consequence,  
the search for a satisfactory notion of multivariate quantile has led to a flourishing literature on this topic despite 
its definition is still a debatable issue (see \cite{chakraborty2003multivariate, hallin2010multivariate, kong2012quantile, koenker2017handbook, Stolfietal, chavas2018multivariate, charlier2020multiple, alfo2020m} and the references therein for relevant studies). Recently, \cite{petrella2019joint} generalized the AL distribution inferential approach of the univariate quantile regression to a multivariate framework by using the Multivariate Asymmetric Laplace (MAL) distribution defined in \cite{kotz2012laplace}. By using the MAL distribution as a likelihood based inferential tool, the authors sidestep the problem of defining the quantiles of a multivariate distribution, and instead implement joint estimation for the univariate quantiles of the conditional distribution of a multivariate response variable given covariates, accounting for possible correlation among the responses.

 \red{In this paper, we generalize} the work of \cite{petrella2019joint} by introducing a Mixed Hidden Markov Model to the longitudinal data setting. \red{From a methodological point of view, we} develop a Quantile Mixed Hidden Markov Model (QMHMM) to jointly estimate the quantiles of the univariate conditional distributions of a multivariate response, accounting for the dependence structure between the outcomes. 
 \red{This is a flexible approach as} time-constant unobserved heterogeneity is described via individual-specific random coefficients while temporal effects are captured through state-specific parameters that evolve over time depending on a hidden Markov chain. 
In order to prevent inconsistent parameter estimates due to misspecification of the random effects distribution, we adopt the Non-Parametric Maximum Likelihood (NPML) approach of \cite{lindsay1983geometry} where the distribution is left unspecified and approximated by a multivariate discrete finite mixture distribution estimated from the data. Within this scheme, our modeling framework reduces to a multivariate finite mixture of HMM quantile regressions.

\red{As in \cite{petrella2019joint}, we propose to estimate the model parameters through Maximum Likelihood (ML) by implementing a suitable Expectation-Maximization (EM) algorithm, which exploits the Gaussian-mixture representation of the MAL distribution where both the hidden Markov chain and the random effects parameters are treated as missing data. Using simulation experiments we assess the validity of the proposed methodology by considering different data generating processes.
\\
 From an applied standpoint, the proposed method is then used to model the quantiles of the conditional distributions of SDQ scores of children in England by using the MCS data. 
 Our approach allows to (i) jointly study the effect of demographic and socio-economic risk factors on internalizing and externalizing disorders in childhood, (ii) investigate whether their effect is more pronounced for children experiencing high levels of disorders (i.e. high quantiles), and (iii) account for the association between SDQ scores and unobserved heterogeneity sources by including into the model time-constant subject-specific random coefficients, without parametric assumptions, and time-dependent random intercepts.
\\
 With respect to existing approaches in the literature, the introduced QMHMM \red{can be thought of as a generalization of the univariate quantile MHMM of \cite{marino2018mixed} to multivariate longitudinal data. Furthermore, our approach} differs substantially from the ones in \cite{kulkarni2019joint} and \cite{alfo2020m}. The authors consider, respectively, univariate quantile and M-quantile (\cite{breckling1988m}) regression models with outcome-specific random effects, where the dependence between responses is captured by time-constant outcome-specific random coefficients. To the best of our knowledge, this is the first attempt 
 to simultaneously estimate multiple conditional quantiles for multivariate longitudinal outcomes, that include both time-constant and time-varying random effects.}

The rest of the paper is organized as follows. In Section \ref{sec:model}, we introduce the proposed QMHMM regression framework. \red{In Section \ref{sec:data} we describe the MCS data. Section \ref{sec:est} illustrates the EM-based ML approach to estimate model parameters together with M-step updates in closed form. In Section \ref{sec:app} we discuss the empirical application while Section \ref{sec:con} summarizes our conclusions.}
 \red{All proofs are presented in Appendix A, while the simulation study is presented in Appendix B.}

\section{\red{The Millennium Cohort Study data}}\label{sec:data}
\red{The MCS study (\url{http://www.cls.ioe.ac.uk}) is a longitudinal, nationally representative study that follows the lives of around 19,000 young people born across UK in 2000-02. The MCS is designed to provide data about children living and growing up in each of the four countries of the UK namely, England, Wales, Scotland and Northern Ireland, especially for sub-groups of children living in advantaged and disadvantaged circumstances, for children of ethnic minorities and those living in Scotland, Wales and Northern Ireland. \green{In addition to} children, it provides data about their families and the broader socio-economic context in which the children live and grow up. 
\\
 The MCS population is defined as all children born between 1 September 2000 and 31 August 2001 (for England and Wales), and between 24 November 2000 and 11 January 2002 (for Scotland and Northern Ireland), living in the UK at age nine months, and whose families were eligible to receive child benefits. In order to meet the principle of adequately representing disadvantaged and ethnic minority children, 
 the population was stratified by UK country. 
 For England, the population was stratified, via the stratification of electoral wards extant on 1 April 1998, into three strata. The first stratum (ethnic stratum) is composed of children living in wards where the proportion of ethnic minorities was at least 30\% in the 1991 Population Census. The second one (disadvantaged stratum) was based on the Child Poverty Index, recording the proportion of children, in a ward whose families received means-tested benefits. The cut-off value to define a disadvantaged ward was 38.4\% receiving such benefits, corresponding to the poorest 25\% wards in England and Wales in 1998.
 The third one (advantaged stratum) consists of all other children. For Wales, Scotland and Northern Ireland, there were just the disadvantaged and advantaged strata. MCS wards were randomly selected within each stratum and country; then a list of all children turning nine months old during the survey window and living in the selected wards was created. The MCS sample members were first surveyed when they were around 9 months of age, and cohort members continue to remain eligible to be surveyed if they remain living in or returning to the UK. Follow-ups have currently been conducted at 3, 5, 7, 11, 14 and 17 years of age. The main data collection from parents at the first survey was by a face-to-face computer assisted interview and self-completion (CAPI and CASI). 
 The full set of questions, typically lasting 70-75 minutes went to a main informant, almost always the mother, and a shorter set of questions, taking around 30 minutes went to the main informant’s partner.
 For further details, please refer to the technical report of \cite{plewis2007millennium} and studies of \cite{connelly2014cohort} and \cite{joshi2016millennium}.}
 This study is widely regarded as the basis of the most reliable estimates of cognitive development problems in young people and it has been \red{thoroughly} investigated in the fields of child psychology and pedagogy: see, among others, the works of \cite{griffiths2011obesity, goodman2011population, tzavidis2016, bell2019relationship, ahn2018associations} and \cite{alfo2020m}. 
 \red{In order to analyze the impact of child and mother personal characteristics, neighbourhood context and family risk factors} on children disorders, the following set of predictors \red{is considered}. ALE 11 measures the number, out of 11 events, of potentially stressful life events experienced by the family between two consecutive sweeps. The events, classified on the basis of the scale proposed by \cite{tiet1998adverse}, are family member died, negative change in financial situation, new step-parent, sibling left home, child got seriously sick or injured, divorce or separation, family moved, parent lost job, new natural sibling, new stepsibling and mother diagnosed with or treated for depression. SED 4 measures the household's socio-economic disadvantage condition by combining information on overcrowding (more than 1.5 people per room excluding the bathroom and kitchen), not owning a home, receipt of means-tested income support and income poverty. Mother's personal characteristics and distress psychological indicators are included such as maternal education (no qualification (baseline), university degree or General Certificate of Secondary Education (GCSE)) and maternal depression, Kessm, measured by the Kessler score \red{and diagnosed by a doctor}. Furthermore, child’s age in years, centered around the mean, age year scal, the quadratic effect of child’s age, age2 year scal, ethnicity (non-white (baseline) or white) and gender (female (baseline) or male) were included in the model. Finally, three explanatory variables evaluate the area characteristics. Imdscore is a time varying variable which measures neighbourhood deprivation by the index of multiple-deprivation score. A design variable which allows for the stratification of the MCS sampling design: the stratification variable of the MCS consists of three categories, namely the advantaged stratum (baseline category), the ethnic stratum, eth stratum, and the disadvantaged stratum, dis stratum. \red{To allow for comparison,} the considered \red{predictors} are the same as \green{those} in \cite{tzavidis2016}.\\ 

\red{The data that we use in this paper are} SDQ internalizing, SDQ$_{Int}$, and SDQ externalizing, SDQ$_{Ext}$, scores recorded on children who were observed at all measurement occasions, i.e. the considered sample consists of $N = 5342$ units and $T_i = T = 3$, for all $i = 1,\dots,N$. \red{
As is customary in child psychology, SDQ scores are treated as though they are continuous variables. Table \ref{tab:summary} presents the main descriptive statistics of continuous and categorical variables considered in the sample. The average values of ALE 11, SED 4 and Kessm are 1.405, 0.555 and 2.597  but, there are cases with much higher scores as demonstrated by their maximum values. 43\% of children have mothers who hold a degree and 48\% have mothers with GCSE or other qualification. Around 50\% of children are males and, in relation to ethnicity, 89\% of members are white. The sample also includes 8.2\% and 38\% of families from the ethnic and disadvantaged strata respectively. Furthermore, the empirical correlation between SDQ$_{Int}$ and SDQ$_{Ext}$ equals to 0.369. As expected, internalizing and externalizing problems are positively correlated, justifying the joint modeling approach we adopt in this paper.} 

\begin{table}[!h]
\centering
 \smallskip 
 \resizebox{0.9\columnwidth}{!}{
\begin{tabular}{lcccccc}
\toprule
Variable & Minimum & 1-st quartile & Median & Mean$^\dagger$ & 3-rd quartile & Maximum\\
\hline\\
SDQ$_{Int}$ & 0 & 1 & 2 & 2.493 & 4 & 19 \\ 
SDQ$_{Ext}$ & 0 & 2 & 5 & 5.144 & 7 & 20 \\
ALE 11 & 0 & 1 & 1 & 1.405 & 2 & 7 \\ 
SED 4 & 0 & 0 & 0 & 0.555 & 1 & 4 \\ 
Kessm & 0 & 0 & 2 & 2.597 & 4 & 24 \\ 
Degree & & & & 43.429 & & \\ 
GCSE & & & & 47.997 & & \\ 
White & & & & 89.012 & & \\ 
Male & & & & 50.337 & & \\ 
IMD & 1 & 3 & 6 & 5.619 & 8 & 10 \\ 
Eth stratum & & & & 8.199 & & \\ 
Dis stratum & & & & 38.132 & & \\ 
\bottomrule
\end{tabular}}
\caption{Summary statistics for the MCS data. $^\dagger$ means for dummy variables are reported in \%.}
\label{tab:summary}
\end{table}

\red{This exploratory analysis and all the preliminary considerations raised in the introduction \green{suggest that} a quantile Mixed Hidden Markov Model that accounts for time-constant and time-dependent unobserved heterogeneity sources \green{offers an approach to modeling the data}. In order to account for the data features and provide correct statistical inferences, the modeling approach we propose defines a joint quantile regression for multivariate longitudinal data to study the effects of \green{environmental}, parental, and child factors across different quantiles of the SDQ distributions. This can offer a more complete picture of the determinants of children’s problems useful 
 for clinicians and educationalists, in order to design programs preventing the onset of psychopathologies and addressing the incidence of disorders in early childhood.} 

\section{Methodology}\label{sec:model}
Let $\mathbf{Y}_{it} = (Y_{it}^{(1)}, \dots, Y_{it}^{(p)})$ be a continuous $p$-variate response variable vector and $\mathbf{X}_{it} = (X_{it}^{(1)}, \dots, X_{it}^{(k)})$ be a $k$-dimensional vector of explanatory variables for subject $i = 1, \dots, N$ and time occasion $t = 1, \dots, T_i$. \red{Let} $\tau = (\tau_1, \dots, \tau_p)$ denote $p$ quantile indexes with $\tau_j \in (0,1)$, for $j = 1, \dots, p$ and let $ S_{it} (\tau)$, $i= 1, \dots, N$, $t = 1, \dots, T_i$ be a homogeneous, first-order, \red{aperiodic and irreducible} hidden Markov chain defined over a discrete states space $\mathcal{S} = \{1, \dots, M \}$ with initial and transition probabilities denoted by $\bs{q} = (q_1, \dots, q_M)$ and $\bs{Q} = \{q_{jk}\}$ over $\mathcal{S}\times\mathcal{S}$ common to all subjects, respectively. Finally, let $\mathbf{b}_i (\tau)$ be a time-constant, subject-specific, random effects matrix having distribution \red{$f_\bs{b} (\cdot \mid \mathbf{X}_{it}, \tau)$ with support $\mathcal{B}$,} 
 where 
$\mathbb{E} (\mathbf{b}_i (\tau)) = 0$ is used for parameter identifiability. We assume that the $\tau_j$-th quantile of each of the $j$-th components of $\mathbf{Y}_{it}$ can be modeled as a function of explanatory variables. 
Let $\boldsymbol{\beta} (\tau) = (\boldsymbol{\beta}_1 (\tau), \dots, \boldsymbol{\beta}_p (\tau))$ be the $k \times p$ matrix of unknown \red{quantile} regression coefficients. 
Then, the QMHMM is defined as follows:
\begin{equation}\label{eq:mhmm}
\mathbf{Y}_{it} = \mathbf{X}_{it} \boldsymbol{\beta} (\tau) + \mathbf{Z}_{it} \mathbf{b}_i (\tau) + \mathbf{W}_{it} \boldsymbol{\alpha}_{S_{it}} (\tau) + \epsilon_{{it}} (\tau)
\end{equation} 
where $\mathbf{Z}_{it}$ is a subset of $\mathbf{X}_{it}$, $\mathbf{W}_{it}$ is a further subset of $\mathbf{X}_{it}$ whose effects are assumed to vary over time, $\epsilon_{{it}} (\tau)$ denotes a $p$-dimensional vector of error terms with univariate component-wise quantiles (at fixed levels $\tau_1,\dots, \tau_p$, respectively) equal to zero and where the coefficients matrix $\boldsymbol{\alpha}_{S_{it}} (\tau)$ evolves over time according to the hidden Markov chain, $S_{it} (\tau)$, and takes one of the values in the set $\{ \boldsymbol{\alpha}_1 (\tau), \dots, \boldsymbol{\alpha}_M (\tau) \}$. \red{In particular, the parameters $\mathbf{b}_i (\tau), i = 1,\dots,N$, and $\{ \boldsymbol{\alpha}_1 (\tau), \dots, \boldsymbol{\alpha}_M (\tau) \}$ are designed to account for within-individual dependence by considering unobserved time-constant and time-varying sources of unobserved heterogeneity, respectively.}
 
 Our objective is to provide joint estimation of the $p$ quantiles of the univariate conditional distributions of $\mathbf{Y}_{it}$ taking into account for potential correlation among the dependent variables. The QMHMM framework is based on the following central assumptions, \red{which are standard in mixed effects models. The random effects $\mathbf{b}_i (\tau)$ are independent of the hidden Markov chain, $S_{it} (\tau)$, as they are meant to capture different unobserved characteristics, and furthermore, it is assumed that the covariates $\mathbf{X}_{it}$ are uncorrelated with $\mathbf{b}_i (\tau)$, that is $f_\bs{b} (\cdot \mid \mathbf{X}_{it}, \tau) = f_\bs{b} (\cdot \mid \tau)$. Regarding the longitudinal responses, they must satisfy the contemporary dependence and conditional independence conditions. The former states that for the $i$-th subject at time $t$, the distribution of $\bs Y_{it}$, given the state variables $(S_{i1} (\tau), ..., S_{iT_i} (\tau))$ and the time-constant individual-specific random effects $\mathbf{b}_i (\tau)$, depends only on the current state $S_{it} (\tau)$; the latter entails that the responses $(\bs Y_{i1}, ..., \bs Y_{iT_i})$ are conditionally independent, given the hidden state occupied at time $t$ by $S_{it} (\tau)$ and the individual-specific random coefficients $\mathbf{b}_i (\tau)$.}
 These assumptions imply that the following equality holds:
\begin{equation}\label{eq:lia}
f_{\bs{Y}} (\bs{y}_{it} \mid \bs{y}_{i1:t-1}, \bs{x}_{i1:t-1}, s_{i1:t}, \bs{b}_i, \tau) = f_{\bs{Y}} (\bs{y}_{it} \mid \bs{x}_{it}, s_{it}, \bs{b}_i, \tau)
\end{equation} 
where $\mathbf{y}_{i1:t-1}$ and $\mathbf{x}_{i1:t-1}$ represent the history of the responses and the observed covariates for the $i$-th subject up to time $t-1$, respectively, and $s_{i1:t}$ is the individual sequence of states up to time $t$.

Generalizing the approach of \cite{petrella2019joint}, for the model in \eqref{eq:mhmm} we consider the MAL distribution, $\mathcal{MAL} (\boldsymbol{\mu}, \bs{D} \bs{\tilde \xi}, \bs{D} \bs \Sigma \bs{D})$, (see \cite{kotz2012laplace}) \red{having density function}:
\begin{equation}\label{eq:MALdensity}
f_{\bs{Y}} (\bs y_{it} \mid \bs{x}_{it}, s_{it}, \bs{b}_i, \tau) =  \frac{ 2 
\exp{\left\{(\bs y_{it}- \boldsymbol{\mu}_{it})' \bs D^{-1} \bs \Sigma^{-1} \bs{\tilde \xi} \right\}}}
{(2\pi)^{p/2} |\bs D \bs \Sigma \bs D|^{1/2}   } \left( \frac{\tilde m_{it}}{2+\tilde d}\right)^{\nu/2}K_{\nu}\left(  \sqrt{(2+\tilde d)\tilde m_{it}} \right), 
\end{equation} 
where the location parameter $\boldsymbol{\mu}_{it}$ is defined by the \textcolor{black}{Mixed Hidden Markov} Model:
\begin{equation}\label{eq:lm}
\boldsymbol{\mu}_{it} = \boldsymbol{\mu} (s_{it}, \bs{b}_i, \tau) = \mathbf{X}_{it} \boldsymbol{\beta} (\tau) + \mathbf{Z}_{it} \mathbf{b}_i (\tau) + \mathbf{W}_{it} \boldsymbol{\alpha}_{s_{it}} (\tau),
\end{equation}
$\bs D \bs {\tilde\xi}$ is the skew parameter with $\bs D= \diag [d_{1} ,\dots, d_{p}]$, $d_{j}>0$ and $ \bs {\tilde\xi}= [\tilde \xi_1, \tilde \xi_2,\dots,\tilde  \xi_p]'$ having generic element  $\tilde \xi_j= \frac{1- 2 \tau_j}{\tau_j(1 - \tau_j)}$, $j=1,\dots, p$. $\bs \Sigma$ is a $p \times p$ positive definite matrix such that $\bs \Sigma = \bs \Lambda \bs \Psi \bs \Lambda$, with $\bs \Psi $ being \red{an unstructured correlation matrix of dimension $p$} and $\bs \Lambda= \diag[\sigma_1,\dots, \sigma_p]$, with $\sigma_j^2= \frac{2}{\tau_j (1 - \tau_j)}$, $j=1,\dots, p$. Moreover, $\tilde m_{it}= (\bs{y}_{it} - \boldsymbol{\mu}_{it})' (\bs D \bs \Sigma \bs D)^{-1}(\bs{y}_{it} -  \boldsymbol{\mu}_{it})$, $\tilde d =\bs {\tilde\xi}' \bs \Sigma^{-1} \bs{\tilde \xi}$, and  $K_{\nu}(\cdot)$ denotes the modified Bessel function of the third kind with index parameter $\nu= (2-p)/2$. 

One of the key benefits of the MAL distribution is that, using \eqref{eq:mhmm} and \eqref{eq:MALdensity}, and following \cite{kotz2012laplace}, \red{$\bs Y \sim \mathcal{MAL}  (\boldsymbol{\mu}, \bs{D} \bs{\tilde \xi}, \bs{D} \bs \Sigma \bs{D})$} can be written as a location-scale mixture, having the following representation:
\begin{equation}\label{eq:mixtureALD}
\mathbf{Y} = \boldsymbol{\mu} + \bs D \bs {\tilde \xi} \tilde C + \sqrt{\tilde C}  \bs D \bs {\Sigma}^{1/2}\bs Z
\end{equation}
where $\bs Z \sim {\cal N}_p(\bs 0_p, \bs I_p)$ denotes a $p$-variate standard Normal distribution and $\tilde C \sim \mbox{Exp}(1)$ has a standard exponential distribution, with  $\bs Z$ being independent of $\tilde C$. In particular, the constraints imposed on $\bs {\tilde\xi}$ and $\bs \Lambda$ represent necessary conditions for model identifiability for any fixed quantile level $\tau_1, \dots, \tau_p$ and guarantee that $\bs{\mu}^{(j)}_{it}$ is the $\tau_j$-th conditional quantile function of $Y_{it}^{(j)}$ given $S_{it} (\tau)$ and $\bs{b}_i$, for $j = 1,\dots, p$. 
 \red{As shown in \cite{petrella2019joint}, \red{using this approach} we are able to conduct inference on the quantiles of the univariate conditional distributions of $\bs Y_{it}$ \red{simultaneously}, taking into account the possible correlation \red{between the outcomes}. For a given quantile level $\tau$, following \cite{kotz2012laplace} and by simple calculations it is possible to show that the covariance matrix of $\bs Y$ can be written as:
\begin{equation}\label{eq:corr}
\bs S = \bs D(\bs {\tilde\xi} \bs {\tilde\xi}' + \bs{ \Lambda} \bs \Psi \bs{ \Lambda}) \bs D,
\end{equation} 
where the off-diagonal elements of $\bs S$ provide an indirect measure of association between the outcomes.}

 Two remarks are \red{also} noteworthy regarding the methodology introduced above. First, our model \textcolor{black}{can be thought of as an extension to multivariate longitudinal data of: (i) the Linear Quantile Hidden Markov Model by \cite{farcomeni2012quantile} when $\mathbf{W}_{it} = \mathbf{1}$ and $\mathbf{b}_i (\tau) = \mathbf{0}$ for all $i = 1, \dots, N$ and $t = 1, \dots, T_i$; (ii) the Linear Quantile Mixed Model (LQMM) proposed in \cite{geraci2014linear} when there is only one state of the hidden Markov chain, i.e. $M=1$.}
 Second, the proposed approach differs substantially from the \red{ones by \cite{kulkarni2019joint} and \cite{alfo2020m}. In the former, the authors consider univariate quantile regression models where the dependence across time and responses is captured by time-constant outcome-specific normally distributed random coefficients. In the latter, the proposed method targets a different set of location parameters, i.e. the M-quantiles (\cite{breckling1988m}) of the distribution of the dependent variables, which are more difficult to interpret than quantiles. The authors then define univariate M-quantile (\cite{breckling1988m}) regression models with outcome-specific random effects, where dependence between outcomes for each unit is introduced by assuming correlated, subject-specific random effects in the univariate models.}


Estimation of model parameters can be pursued using a ML approach. To ease the notation, unless specified otherwise, hereinafter we omit the quantile levels vector $\tau$, yet all model parameters are allowed to depend on the $p$ quantile indexes. Thus, let us denote by $\bs \Phi_{\bs{\tau}} = (\boldsymbol{\beta}, \bs{D}, \bs \Psi, \boldsymbol{\alpha}_1, \dots, \boldsymbol{\alpha}_M, \bs{q}, \bs{Q})$ the set of model parameters. \red{Given the modeling assumptions introduced so far, the} observed \red{data} likelihood is defined by:
\begin{equation}\label{eq:ollk}
L(\bs \Phi_{\bs{\tau}}) = \prod_{i=1}^N \int_{\mathcal{B}} \Bigg\{ \sum_{\mathcal{S}^{T_i}} \Bigg[ \prod_{t=1}^{T_i} f_{\bs{Y}} (\bs y_{it} \mid \bs{x}_{it}, s_{it}, \bs{b}_i)  \Bigg] q_{s_{i1}} \prod_{t=2}^{T_i} q_{s_{it-1}s_{it}} \Bigg\} f_{\bf b} ({\bf b}_i) \textnormal{d} {\bf b}_i.
\end{equation}
The maximization of the likelihood in \eqref{eq:ollk} generally may prove to be excessively cumbersome because it involves a multidimensional integral over the random coefficients distribution $f_{\bf b} (\cdot)$ and a summation over $M^{T_i}$ terms for each unit. In addition, the choice of an appropriate distribution for the random effects is not straightforward. Ideally $f_{\bf b} (\cdot)$ should be data driven and resistant to misspecification \citep{marino2015linear}, otherwise an incorrect distributional assumption for the random effects has unfavorable influence on statistical inferences (see \cite{agresti2004examples, maruotti2011mixed} and \cite{neuhaus2013estimation}). In the next section, we discuss how \red{we specify the random effects distribution and how} we may avoid evaluating the integral in \eqref{eq:ollk} for ML estimation.

\subsection{Specification of the random coefficients distribution}\label{sub:ran}
In the literature, typically the Gaussian distribution is a convenient choice for $f_{\bf b} (\cdot)$ from a computational point of view. In this case, we may approximate the integral in \eqref{eq:ollk} using Gaussian quadrature or adaptive Gaussian quadrature schemes (see \cite{rabe2005maximum, pinheiro2006efficient} and \cite{crowther2014multilevel}). A disadvantage of such approaches lies in the required computational effort, which is exponentially increasing with the number of the random parameters. For this reason, potential alternatives \red{proposed the use} of simulation methods such as Monte Carlo and simulated ML approaches (\cite{mcculloch1997maximum}). However, for samples of finite size and short individual sequences, these methods may not provide a good approximation of the true mixing distribution (\cite{alfo2017finite}). As a robust alternative to the Gaussian choice, the multivariate Symmetric Laplace or multivariate Student t distributions have been considered by \cite{geraci2014linear} and \cite{farcomeni2015longitudinal}. However, a parametric assumption on the distribution of the random coefficients could be rather restrictive and misspecification of the mixing distribution can lead to biased parameter estimates (see \cite{Alfo2010}). 
 Following \cite{marino2018mixed}, in this work we exploit the approach based on the Non-Parametric Maximum Likelihood (NPML) estimation of \cite{laird1978nonparametric} \red{and extend it to the multivariate context}. In particular, we do not parametrically specify $f_{\bf b} (\cdot)$ but we approximate it by using a discrete distribution \red{defined on $G < N$ multivariate locations, $\textbf{b}_{g} (\tau)$, with associated probabilities defined by:
\begin{equation}\label{eq:pig}
\pi_g (\tau) = \textnormal{Pr}({\bf b}_i (\tau) = {\bf b}_g (\tau)),
\end{equation}
with $\pi_g \geq 0$, $\forall \, \, g = 1, \dots , G$ and $\sum_{g=1}^G \pi_g = 1$.
 More concisely, we can write:
\begin{equation}\label{eq:prox}
{\bf b}_i (\tau) \sim \sum_{g=1}^G \pi_g (\tau) \delta_{{\bf b}_g} (\tau),
\end{equation}
where $\delta_\theta$ is a one-point distribution putting a unit mass at $\theta$. \red{With this approach,} the parametric problem is thus converted to a semiparametric one, where $\textbf{b}_{g} (\tau)$ and $\pi_g (\tau)$ \red{define} the discrete probability distribution of the random effects defined on $G$ distinct support points. In this context, time-constant unobserved heterogeneity in the data is represented by a finite mixture with unknown proportions $\pi_g (\tau)$ and locations $\textbf{b}_{g} (\tau)$ common to all subjects in the $g$-th group.}
 \red{Since locations and masses are completely free to vary over the corresponding support, 
 this is a flexible method that can readily accommodate a wide range of shapes, including fat-tailed and asymmetric distributions, and it is more robust against deviations from model assumptions.
 For the interested reader, a detailed survey \red{about} this method can be found in \cite{Aitkin1998, Alfo2000, Aitkin2003, alfo2017finite, alfo2020m} and \cite{maruotti2020two}, for example.
 \\
 In this setting, the observed data likelihood in \eqref{eq:ollk} reduces to:}
\begin{equation}\label{eq:llk}
L({\bf \Phi}_{\bs{\tau}}) = \prod_{i=1}^N \sum_{g=1}^G \Bigg\{ \sum_{\mathcal{S}^{T_i}} \Bigg[ \prod_{t=1}^{T_i} f_{\bs{Y}} (\bs y_{it} \mid \bs x_{it}, s_{it}, \bs{b}_g)  \Bigg] q_{s_{i1}} \prod_{t=2}^{T_i} q_{s_{it-1}s_{it}} \Bigg\} \pi_g,
\end{equation}
where ${\bf \Phi}_{\bs{\tau}} = (\boldsymbol{\beta}, \bs{D}, \bs \Psi, \bs{b}_1, \dots, \bs{b}_G, \pi_1, \dots, \pi_G, \boldsymbol{\alpha}_1, \dots, \boldsymbol{\alpha}_M, \bs{q}, \bs{Q})$ denotes the vector of model parameters and $f_{\bs{Y}} (\bs y_{it} \mid \bs x_{it}, s_{it}, \bs{b}_g)$ represents the response distribution of unit $i$ being in the state $s_{it}$ at time $t$ and belonging to the $g$-th component of the finite mixture, which is assumed to \red{follow the} MAL \red{as} in \eqref{eq:MALdensity} with location parameter given by:
\begin{equation}\label{eq:lmf}
\boldsymbol{\mu}_{it} = \boldsymbol{\mu} (s_{it}, \bs{b}_g, \tau) = \mathbf{X}_{it} \boldsymbol{\beta} (\tau) + \mathbf{Z}_{it} \mathbf{b}_g (\tau) + \mathbf{W}_{it} \boldsymbol{\alpha}_{s_{it}} (\tau).
\end{equation}
By looking at the likelihood in \eqref{eq:llk}, one can see that it resembles the likelihood of a finite mixture of HMM models with $G$ homogeneous sub-populations where 
the presence of latent time-constant heterogeneity is described by discrete multivariate random effects. \red{From the estimation perspective, locations $\bs{b}_g$ and corresponding probabilities $\pi_g$ are unknown parameters which need to be estimated along with other model parameters. The number of mixture components $G$ is unknown, and it is usually treated as fixed and estimated via penalized likelihood criteria (see e.g. \cite{bohning1999computer}). Furthermore, as an important by-product, the computational complexity of the likelihood evaluation in \eqref{eq:llk} is of linear order with respect to $G$, which greatly facilitates the implementation of EM-type algorithm, 
 as is described in the following section.}

\section{Maximum Likelihood Estimation and Inference}\label{sec:est}
As \red{mentioned} in the previous sections, the MAL density represents a convenient tool to jointly model the univariate quantiles of the conditional distribution of a multivariate response variable in a quantile regression framework. In this section we introduce a ML approach to estimate and make inference on model parameters \red{and} \red{build} a suitable EM algorithm \citep{dempster1977maximum}. We will show that the M-step update of all model parameters can be easily obtained in closed form, hence reducing the computational burden of the algorithm compared to direct maximization of the likelihood in \eqref{eq:llk}. Specifically, we derive \red{the} EM algorithm by exploiting the Gaussian location-scale mixture representation in \eqref{eq:mixtureALD} of the MAL distribution under the constraints on $\bs {\tilde\xi}$ and $\bs \Lambda$. 
%


\subsection{The EM algorithm}\label{sub:EM}
The EM algorithm alternates between an expectation (E) step, which defines the expectation of the complete log-likelihood evaluated \red{of} the current \red{parameters} estimates, and a maximization (M) step, which computes parameter estimates by maximizing the expected complete log-likelihood obtained in the E-step. The complete log-likelihood, the expected complete log-likelihood function and the optimal parameter estimators are given below  in the following propositions. 

Given the representation in \eqref{eq:llk}, let us denote by $w_{ig}$ the indicator variable that is equal to $1$ if the $i$-th unit belongs to the $g$-th component of the finite mixture, and 0 otherwise. Similarly, let $u_{itj}$ be equal to $1$ if unit $i$ is in state $j$ at time $t$ and 0 otherwise; let $v_{itjk}$ be equal to 1 if unit $i$ is in state $j$ at time $t-1$ and in state $k$ at time $t$, and 0 otherwise. Finally, we denote by $z_{itjg}$ the indicator of the $i$-th individual being in state $j$ at time $t$ and coming from the $g$-th component of the mixture. The expected \red{complete data log-likelihood} is presented in the following proposition.

\begin{proposition}\label{prop:completeloglik}
For any fixed $\bs \tau=[\tau_1, \tau_2,\dots,\tau_p]$, $G$ mixture components and $M$ hidden states, the complete data log-likelihood function \red{is proportional to}:
\begin{equation}\label{eq:OF}
\begin{split}
\ell_c({\bf \Phi}_{\bs{\tau}}) & \propto \sum_{i=1}^N \Bigg\{ \sum_{g=1}^G w_{ig} \log \pi_g + \sum_{j=1}^M u_{i1j} \log q_j + \sum_{t=2}^{T_i} \sum_{j=1}^M \sum_{k=1}^M v_{itjk} \log q_{jk} \\
& - \frac{1}{2} T_i \log \mid \bs D {\bs \Sigma} \bs D \mid + \sum_{t=1}^{T_i} \sum_{j=1}^M \sum_{g=1}^G z_{itjg} (\bs Y_{it} - \boldsymbol{\mu}_{it})' \bs D^{-1} {\bs \Sigma}^{-1}\tilde {\bs \xi} \\
& - \frac{1}{2} \sum_{t=1}^{T_i} \sum_{j=1}^M \sum_{g=1}^G z_{itjg} \frac{1}{\tilde C_{itjg}} (\bs Y_{it} - \boldsymbol{\mu}_{it})' (\bs D {\bs \Sigma} \bs D)^{-1} (\bs Y_{it} - \boldsymbol{\mu}_{it}) \\
& - \frac{1}{2} \tilde{{\bs \xi}}' {\bs \Sigma^{-1}} \tilde {\bs \xi} \sum_{t=1}^{T_i} \sum_{j=1}^M \sum_{g=1}^G z_{itjg} \tilde C_{itjg} \Bigg\},
\end{split}
\end{equation}
where $\tilde C_{itjg}$ is a latent variable that follows an exponential distribution with parameter $1$.
\end{proposition}
In the E-step of the algorithm, the presence of the unobserved indicator variables $w_{ig}, u_{itj}, v_{itjk}$ and $z_{itjg}$ is handled by taking their conditional expectation given the observed data and the current parameter estimates. Calculation of such quantities may be addressed via an adaptation of the forward and backward variables; see \cite{welch2003hidden}. Similarly, the conditional expectations of $\frac{1}{\tilde C_{itjg}}$ and $\tilde C_{itjg}$ are considered.

For the implementation of the algorithm, forward and backward variables are defined for the longitudinal measures. We define the probability of observing the partial sequence ending up in state $j$ at time $t$, given the $g$-th component, as:
\begin{equation}
a_{it} (j,g) = f(\bs{y}_{i1:t}, S_{it} = j \mid \bs {b}_g) \qquad \textnormal{and} \qquad a_{i1} (j,g) = q_j f(\bs{y}_{i1} \mid S_{i1} = j, \bs {b}_g).
\end{equation}
The quantity $a_{it} (j,g)$ can be rewritten using the following recurrence relationship:
\begin{equation}
a_{it} (j,g) = \sum_{h=1}^M a_{it-1} (h,g) q_{hj} f(\bs{y}_{it} \mid S_{it} = j, \bs {b}_g).
\end{equation}
Backward variables are defined as the probability of the longitudinal sequence from time $t+1$ to the last available observation $T_i$, conditional on being in state $j$ at time $t$, given the $g$-th component:
\begin{equation}
b_{it} (j,g) = f(\bs{y}_{it+1:T_i} \mid S_{it} = j, \bs {b}_g) \qquad \textnormal{and} \qquad b_{iT_i} (j,g) = 1.
\end{equation}
Accordingly, the backward variable $b_{it} (j,g)$ can be rewritten as:
\begin{equation}
b_{it} (j,g) = \sum_{k=1}^M b_{it+1} (k,g) q_{jk} f(\bs{y}_{it+1} \mid S_{it+1} = k, \bs {b}_g).
\end{equation}
Finally, the expected values of $w_{ig}, u_{itj}, v_{itjk}$ and $z_{itjg}$ can be computed as:
\begin{equation}\label{eq:posterior}
\begin{split}
\hat w_{ig} & = \frac{\pi_g \sum_{j=1}^M a_{iT_i} (j,g)}{\sum_{g=1}^G \pi_g \sum_{j=1}^M a_{iT_i} (j,g)}, \\
\hat z_{itjg} & = \frac{a_{it} (j,g) b_{it} (j,g) \pi_g}{\sum_{g=1}^G \sum_{j=1}^M a_{it} (j,g) b_{it} (j,g) \pi_g}, \\
\hat u_{itj} & = \sum_{g=1}^G \hat z_{itjg}, \\
\hat v_{itjk} & = \frac{\sum_{g=1}^G a_{it-1} (j,g) q_{jk} f(\bs{y}_{it} \mid S_{it} = k, \bs {b}_g) b_{it} (k,g) \pi_g}{\sum_{g=1}^G \sum_{j=1}^M \sum_{k=1}^M a_{it-1} (j,g) q_{jk} f(\bs{y}_{it} \mid S_{it} = k, \bs {b}_g) b_{it} (k,g) \pi_g}.
\end{split}
\end{equation}
By substituting the corresponding posterior expectations in \eqref{eq:posterior} into the complete data likelihood in \eqref{eq:OF}, the expected complete \red{data} log-likelihood function is provided in the following proposition.
\begin{proposition}\label{prop:expectedloglik}
For any fixed $\bs \tau=[\tau_1, \tau_2,\dots,\tau_p]$, $G$ mixture components and $M$ hidden states, the expected complete data log-likelihood function \red{is proportional to}:
\begin{equation}\label{eq:OF2}
\begin{split}
\mathcal{O}({\bf \Phi}_{\bs{\tau}}) & \propto \sum_{i=1}^N \Bigg\{ \sum_{g=1}^G \hat w_{ig} \log \pi_g + \sum_{j=1}^M \hat u_{i1j} \log q_j + \sum_{t=2}^{T_i} \sum_{j=1}^M \sum_{k=1}^M \hat v_{itjk} \log q_{jk} \\
& - \frac{1}{2} T_i \log \mid \bs D {\bs \Sigma} \bs D \mid + \sum_{t=1}^{T_i} \sum_{j=1}^M \sum_{g=1}^G \hat z_{itjg} (\bs Y_{it} - \boldsymbol{\mu}_{it})' \bs D^{-1} {\bs \Sigma}^{-1}\tilde {\bs \xi} \\
& - \frac{1}{2} \sum_{t=1}^{T_i} \sum_{j=1}^M \sum_{g=1}^G \hat z_{itjg} \hat{\tilde z}_{itjg} (\bs Y_{it} - \boldsymbol{\mu}_{it})' (\bs D {\bs \Sigma} \bs D)^{-1} (\bs Y_{it} - \boldsymbol{\mu}_{it}) \\
& - \frac{1}{2} \tilde{{\bs \xi}}' {\bs \Sigma^{-1}} \tilde {\bs \xi} \sum_{t=1}^{T_i} \sum_{j=1}^M \sum_{g=1}^G \hat z_{itjg} \hat{\tilde c}_{itjg} \Bigg\},
\end{split}
\end{equation}
where
\begin{equation}\label{eq:wi}
\hat{\tilde c}_{itjg} =  \left( \frac{\tilde {m}_{itjg}}{2+ \tilde d} \right)^{\frac{1}{2}} \frac{K_{\nu +1}\left( \sqrt{(2+ \tilde d) {\tilde {m}}_{itjg}} \right)}{K_{\nu}\left(   \sqrt{(2+ \tilde d)  {\tilde {m}}_{itjg}}\right)}, \qquad 
\hat{\tilde z}_{itjg} = \left( \frac{2+ {\tilde d}}{{\tilde {m}}_{itjg}} \right)^{\frac{1}{2}}  \frac{K_{\nu +1}  \left( \sqrt{(2+ {\tilde d}) {\tilde {m}}_{itjg}} \right)}{K_{\nu}  \left( \sqrt{(2+ {\tilde d}) {{\tilde {m}}_{itjg}}} \right)} - \frac{2 \nu}{{\tilde {m}}_{itjg}}, 
\end{equation}
with
\begin{equation}
\tilde {m}_{itjg} = (\bs y_{it} - \boldsymbol{\mu}_{it})' (\bs D {{\bs \Sigma}} \bs D)^{-1} (\bs y_{it} - \boldsymbol{\mu}_{it}), \qquad 
{ \tilde d} = \tilde {\bs \xi}' {{\bs \Sigma^{-1}} } \tilde{ \bs \xi}.
\end{equation}
\end{proposition}

Therefore, the EM algorithm can be implemented as follows: 

\medskip

\textit{E-step}: At $r$-th iteration of the algorithm, let $\hat{{\bf \Phi}}^{(r-1)}_\tau$ denote the current parameter estimates. Then, conditionally on the observed data and $\hat{{\bf \Phi}}^{(r-1)}_\tau$, calculate the \red{conditional expectations} in \eqref{eq:posterior} and \eqref{eq:wi}. We denote such quantities $\hat w^{(r)}_{ig}, \hat z^{(r)}_{itjg}, \hat u^{(r)}_{itj}, \hat v^{(r)}_{itjk}$, and $\hat{\tilde c}^{(r)}_{itjg}, \hat{\tilde z}^{(r)}_{itjg}$.

\textit{M-step}: Use $\hat w^{(r)}_{ig}, \hat z^{(r)}_{itjg}, \hat u^{(r)}_{itj}, \hat v^{(r)}_{itjk}$, and $\hat{\tilde c}^{(r)}_{itjg}, \hat{\tilde z}^{(r)}_{itjg}$ to maximize $\mathcal{O}({\bf \Phi}_{\bs{\tau}} \mid \hat{{\bf \Phi}}^{(r-1)}_\tau)$ with respect to ${\bf \Phi}_\tau$, and obtain the update parameter estimates. Based on the introduced modeling assumptions, the maximization can be partitioned into orthogonal subproblems, i.e. the maximization with respect to the fixed, hidden Markov chain and discrete mixing distribution parameters can be performed separately. The initial probabilities $q_j$, transition probabilities $q_{jk}$ and mixing proportions $\pi_g$ are estimated by:
\begin{equation}\label{eq:Mupdates}
\hat q^{(r)}_j = \frac{\sum_{i=1}^N \hat u^{(r)}_{i1j}}{N}, \qquad \hat q^{(r)}_{jk} = \frac{\sum_{i=1}^N \sum_{t=1}^{T_i} \hat v^{(r)}_{itjk}}{\sum_{i=1}^N \sum_{t=1}^{T_i} \sum_{k=1}^M \hat v^{(r)}_{itjk}}, \qquad \hat \pi^{(r)}_g = \frac{\sum_{i=1}^N \hat w^{(r)}_{ig}}{N}.
\end{equation}

If we let $S_{it} = j$, this implies that $\boldsymbol{\alpha}_{s_{it}} = \boldsymbol{\alpha}_j$ and the M-step updates of model parameters $\boldsymbol{\beta} , {\bs b}_g , \boldsymbol{\alpha}_j , {\bs{\Sigma}}$ and $ \bs D$, are given in the following proposition.
\begin{proposition}\label{prop:Mupdates}
The values of $\boldsymbol{\beta} , {\bs b}_g , \boldsymbol{\alpha}_j , {\bs{\Sigma}}$ and $ \bs D$ maximizing \eqref{eq:OF2} are:
\begin{equation}\label{eq:focBeta}
\begin{split}
\hat{\boldsymbol{\beta}}^{(r)}  & = ( \sum_{i=1}^N \sum_{t=1}^{T_i} \sum_{g=1}^G \sum_{j=1}^M \hat z^{(r)}_{itjg} \hat{\tilde z}^{(r)}_{itjg} \bs X'_{it} \bs X_{it} )^{-1} ( \sum_{i=1}^N \sum_{t=1}^{T_i} \sum_{g=1}^G \sum_{j=1}^M \hat z^{(r)}_{itjg} \hat{\tilde z}^{(r)}_{itjg} \bs X'_{it} \bs{\tilde{Y}}_{it} \\
& - \sum_{i=1}^N \sum_{t=1}^{T_i} \sum_{g=1}^G \sum_{j=1}^M z^{(r)}_{itjg} \bs X'_{it} \tilde{\bs \xi}' {\bs D}^{(r-1)}),
\end{split}
\end{equation}
where $\bs{\tilde{Y}}_{it} = \bs{Y}_{it} - \mathbf{Z}_{it} \hat{{\bs b}}^{(r-1)}_g  - \mathbf{W}_{it} \hat{\boldsymbol{\alpha}}^{(r-1)}_j$.
\begin{equation}\label{eq:focb}
\begin{split}
\hat{{\bs b}}^{(r)}_g  & = ( \sum_{i=1}^N \sum_{t=1}^{T_i} \sum_{j=1}^M \hat z^{(r)}_{itjg} \hat{\tilde z}^{(r)}_{itjg} \bs Z'_{it} \bs Z_{it} )^{-1} ( \sum_{i=1}^N \sum_{t=1}^{T_i} \sum_{j=1}^M \hat z^{(r)}_{itjg} \hat{\tilde z}^{(r)}_{itjg} \bs Z'_{it} \bs{\tilde{Y}}_{it} - \sum_{i=1}^N \sum_{t=1}^{T_i} \sum_{j=1}^M z^{(r)}_{itjg} \bs Z'_{it} \tilde{\bs \xi}' {\bs D}^{(r-1)}),
\end{split}
\end{equation}
where $\bs{\tilde{Y}}_{it} = \bs{Y}_{it} - \mathbf{X}_{it} \hat{\boldsymbol{\beta}}^{(r)}  - \mathbf{W}_{it} \hat{\boldsymbol{\alpha}}^{(r-1)}_j$.
\begin{equation}\label{eq:focalpha}
\begin{split}
\hat{\boldsymbol{\alpha}}^{(r)}_j & = ( \sum_{i=1}^N \sum_{t=1}^{T_i} \sum_{g=1}^G \hat z^{(r)}_{itjg} \hat{\tilde z}^{(r)}_{itjg} \bs W'_{it} \bs W_{it} )^{-1} ( \sum_{i=1}^N \sum_{t=1}^{T_i} \sum_{g=1}^G \hat z^{(r)}_{itjg} \hat{\tilde z}^{(r)}_{itjg} \bs W'_{it} \bs{\tilde{Y}}_{it} - \sum_{i=1}^N \sum_{t=1}^{T_i} \sum_{g=1}^G z^{(r)}_{itjg} \bs W'_{it} \tilde{\bs \xi}' {\bs D}^{(r-1)}),
\end{split}
\end{equation}
where $\bs{\tilde{Y}}_{it} = \bs{Y}_{it} - \mathbf{X}_{it} \hat{\boldsymbol{\beta}}^{(r)}  - \mathbf{Z}_{it} \hat{{\bs b}}^{(r)}_g$.
\begin{equation}\label{eq:focSigma}
\begin{split}
\hat{\bs \Sigma}^{(r)} & = \frac{1}{\sum_{i=1}^N T_i} \sum_{i=1}^N \sum_{t=1}^{T_i} \sum_{g=1}^G \sum_{j=1}^M \hat z^{(r)}_{itjg} \hat{\tilde z}^{(r)}_{itjg} \hat{\bs D}^{-1}{}^{(r-1)} (\bs Y_{it} - \hat{\boldsymbol{\mu}}^{(r)}_{it})' (\bs Y_{it} - \hat{\boldsymbol{\mu}}^{(r)}_{it}) \hat{\bs D}^{-1}{}^{(r-1)} \\
& + \frac{1}{\sum_{i=1}^N T_i} \sum_{i=1}^N \sum_{t=1}^{T_i} \sum_{g=1}^G \sum_{j=1}^M \hat z^{(r)}_{itjg} \hat{\tilde c}^{(r)}_{itjg} \tilde{\bs \xi} \tilde {\bs \xi}'
- \frac{2}{\sum_{i=1}^N T_i} \hat{\bs D}^{-1}{}^{(r-1)} \sum_{i=1}^N \sum_{t=1}^{T_i} \sum_{g=1}^G \sum_{j=1}^M \hat z^{(r)}_{itjg} (\bs Y_{it} - \hat{\boldsymbol{\mu}}^{(r)}_{it})' \tilde {\bs \xi}',
\end{split}
\end{equation}
where $\hat{\boldsymbol{\mu}}^{(r)}_{it} = \mathbf{X}_{it} \hat{\boldsymbol{\beta}}^{(r)}  + \mathbf{Z}_{it} \hat{{\bs b}}^{(r)}_g  + \mathbf{W}_{it} \hat{\boldsymbol{\alpha}}^{(r)}_j$.

\medskip

Finally, the elements $d_j, j = 1,\dots,p$ of the diagonal scale matrix $\bs D$ are estimated by:
\begin{equation}\label{eq:focD}
\hat{d}^{(r)}_j = \frac{1}{\sum_{i=1}^N T_i} \sum_{i=1}^N \sum_{t=1}^{T_i} \sum_{g=1}^G \sum_{k=1}^M \hat{z}^{(r)}_{itkg} \rho_\tau (Y_{it}^{(j)} - \hat{{\mu}}^{(j)}_{it}{}^{(r)}),
\end{equation}
where $\rho_\tau (\cdot)$ is the quantile check function of \cite{koenker1978regression}:
\begin{equation}
\rho_\tau (u) = u (\tau - \boldsymbol{1}(u < 0))
\end{equation}
and $\hat{{\mu}}^{(j)}_{it}{}^{(r)}$ is the $j$-th element of the vector $\hat{\boldsymbol{\mu}}^{(r)}_{it}$.
\end{proposition}
The E- and M-steps are alternated until convergence, that is when $\mid \hat{{\bf \Phi}}^{(r)}_\tau - \hat{{\bf \Phi}}^{(r-1)}_\tau \mid$ is smaller than a predetermined threshold. In this paper, we set this convergence criterion equal to $10^{-6}$. 

 \red{Because both the number of components of the finite mixture and hidden states of the Markov chain are unknown a-priori, we select the optimal value of $G$ and $M$ using the BIC (\cite{schwarz1978estimating}):
\begin{equation}\label{eq:BIC}
BIC_{(G,M)} = - 2 \ell (\bs \Phi_\tau) + \log(N) \nu_f,
\end{equation}
where $\ell (\bs \Phi_\tau)$ is the observed data log-likelihood in \eqref{eq:llk}, $N$ is the number of observed individuals and $\nu_f$ denotes the number of free model parameters in $\bs \Phi_\tau$. 
 Following \cite{marino2018mixed}, to avoid convergence to local maxima and better explore the parameter space, for fixed $\tau, G$ and $M$, we fit the QMHMM model using a multiple random starts strategy with 50 different starting points and retain the solution corresponding to the maximum likelihood value. We then repeat this procedure for a grid of values of $G$ and $M$, and select the best combination of the pair $(G, M)$ corresponding to the lowest BIC value.}
 \red{The validity of the proposed EM algorithm and model selection procedure have been assessed using also a simulation exercise (see Appendix B).}

Standard errors of model parameters \red{are computed} using a non-parametric block bootstrap. That is, by re-sampling individuals with replacement and retaining the corresponding sequence of measurements to preserve the within individual dependence structure (see \cite{geraci2014linear, marino2015linear} and \cite{marino2018mixed} for example). We \red{refit} the model to $H$ bootstrap samples and approximate the standard error of each model parameter with \red{the square root of the variance as follows:}
\begin{equation}\label{eq:stand}
\widehat{\textnormal{Cov}} (\hat{{\bf \Phi}}_\tau) = \sqrt{\frac{1}{H-1} \sum_{h=1}^H (\hat{{\bf \Phi}}^{(h)}_\tau - \bar{{\bf \Phi}}_\tau) (\hat{{\bf \Phi}}^{(h)}_\tau - \bar{{\bf \Phi}}_\tau)'},
\end{equation}
where $\hat{{\bf \Phi}}^{(h)}_\tau$ is the set of parameter estimates for the $h$-th bootstrap sample and $\bar{{\bf \Phi}}_\tau$ denote the mean of \red{the model parameters over bootstrap iterations. The standard errors are given by the diagonal elements of $\widehat{\textnormal{Cov}} (\hat{{\bf \Phi}}_\tau)$.}

\section{\red{Analysis of the Millennium Cohort Study data}}\label{sec:app}
\red{In this section, we analyze internalizing and externalizing data on disorders of children collected in the MCS dataset. \red{We are interested in investigating} the impact of environmental, parental, and child factors across \red{both} the distributions of SDQ scores.}
 \red{In order to account for all the data features described in Section \ref{sec:data}, 
 we consider \red{a bivariate} QMHMM with time-varying random intercepts and constant random slopes specified for age \red{to jointly model internalizing and externalizing disorders}. 
 We fitted the proposed model at quantile levels $\tau = (0.25, 0.25), \tau = (0.50, 0.50)$, $\tau = (0.75, 0.75)$ and $\tau = (0.90, 0.90)$. Considering the 75-th and 90-th percentiles puts emphasis on children with more severe problems generally associated with higher levels of SDQ scores. We estimated the QMHMM for a varying number of hidden states $(M = 2, \dots, 8)$ and mixture components $(G = 2, \dots,8)$ employing the multi-start strategy described in Section \ref{sec:est}, and then selected the optimal value of the pair $(G,M)$ corresponding to the lowest BIC value.
 Following \cite{marino2018mixed}, } to enhance model interpretability and produce meaningful results, we retain only those solutions ensuring $\pi_g > 0.05$ for $g = 1,\dots,G$ and $q_j > 0.05$ for $j = 1,\dots,M$. 

 \red{In addition to the proposed model, we compare our methodology with two well-known univariate alternatives for modeling longitudinal data: (i) the Linear Random Effects Model (LREM) for the mean with time-constant random intercepts; (ii) the LQMM of \cite{geraci2014linear} with time-constant random intercepts, at quantile level $\tau = (0.25, 0.50, 0.75, 0.90)$. Specifically, the two models are estimated on SDQ$_{Int}$ and SDQ$_{Ext}$ scores independently. 
 The reason why we consider the LREM is because it is a popular model for targeting the conditional expectation of the response given the explanatory variables. Whilst it produces efficient results when the normality assumptions hold, the LREM could potentially miss out important information related to other parts of the distribution of the outcome.
 In this case, the conditional mean may not offer the best summary; by contrast, the LQMM with a random-intercept specification has a correlation structure that is simple to estimate while allowing for modeling the entire conditional distribution of the outcome. However, both models are using a univariate approach, which completely disregards the possible dependence between the SDQ scores, and assume time-constant random intercepts.} \green{In contrast, the proposed model allows for the correlation between the responses.} 
 
\subsection{\red{Results}}\label{subsec:res}
 \red{We start by commenting on the QMHMM results. Table \ref{tab:est} reports point estimates of model parameters for the two outcomes and standard errors (in parentheses) based on $B = 1000$ bootstrap re-samples. Parameter estimates are displayed in boldface when significant at the standard 5\% level. As one can see, \red{the model selection procedure described in Section \ref{sub:EM} leads to} an increasing number of mixture components $G$ equal to 3, 5, 5 and 5, and a decreasing number of hidden states $M$ equal to 5, 4, 3 and 3 at quantile levels \red{$(0.25, 0.25), (0.50, 0.50), (0.75, 0.75)$ and $(0.90, 0.90)$}, respectively. The chosen values for $G$ and $M$ confirm the presence of constant and serial latent heterogeneity in the data, and support the exploratory analysis of individual SDQ trajectories in Figure \ref{fig:ind}. \red{Moreover, they allow us to classify children based on the intensity of mental and behavioural problems.}}
 
 \red{The second crucial finding is that the coefficient estimates vary with the quantile level $\tau$ and the effect of the covariates appears to be more pronounced at the right tail of the distribution of the responses. In particular, 
 increasing adverse life events, socio-economic disadvantage, maternal depression and low maternal education are statistically associated with both SDQ scores and their impact increases} when looking at the upper tail compared to the lower tail of the distribution. 
 \red{This indicates that an unstable and fragmented family environment has a greater adverse impact on children facing more problems.} Regarding income, there is evidence that poorer children are more likely to suffer from both physical and mental health problems (\cite{currie2009healthy}), hence the role of family income is likely to be concentrated at low incomes (see \cite{fitzsimons2017poverty}). Moreover, maternal depression has a more pronounced effect at the top end where children display critical levels of adjustment problems than at the bottom end of the distribution (see \cite{kiernan2008economic}). These considerations suggest that low socioeconomic status creates stress within the household, causing poorer child health. \red{In relation to gender, males present lower internalizing problems at low quantiles compared to females, while the estimated effect is statistically significant and positive at $\tau = (0.9, 0.9)$. On the other hand, ethnicity and ethnic stratification variables do not appear to be associated with the responses at the 25-th percentiles.}
  
By looking at the fixed parameter estimates for the SDQ$_{Ext}$, \green{we} observe that, in contrast with internalizing scores, boys present more externalizing problems than girls (\cite{flouri2016prosocial}) and the effect is more exacerbated in the right tail of the distribution. In general, girls are at lower risk of behavioral problems than boys which experience an increased risk for conduct and hyperactivity problems (\cite{carona2014social}). Stressful life events, socio-economic disadvantage, maternal depression and maternal education are all significantly associated with internalizing scores. The effect of the covariates is not uniform across quantiles but it is more apparent as the quantile level increases. 
 Moreover, the impact of such variables is more pronounced, across the distribution, on externalizing than on internalizing scores. This is consistent with other studies on behaviour disorders in child psychopathology claiming that poverty and material deprivation and education are more strongly related with children's externalizing problems compared with internalizing problems (see \cite{costello2003relationships} and \cite{dearing2006within}). 
 Overall, point estimates of regression coefficients are consistent with child development theory, as well as with the results discussed in \cite{tzavidis2016} and \cite{alfo2020m}. 
 
\begin{figure}[h]
\center
\includegraphics[width=1\linewidth, height=6.25cm, keepaspectratio]{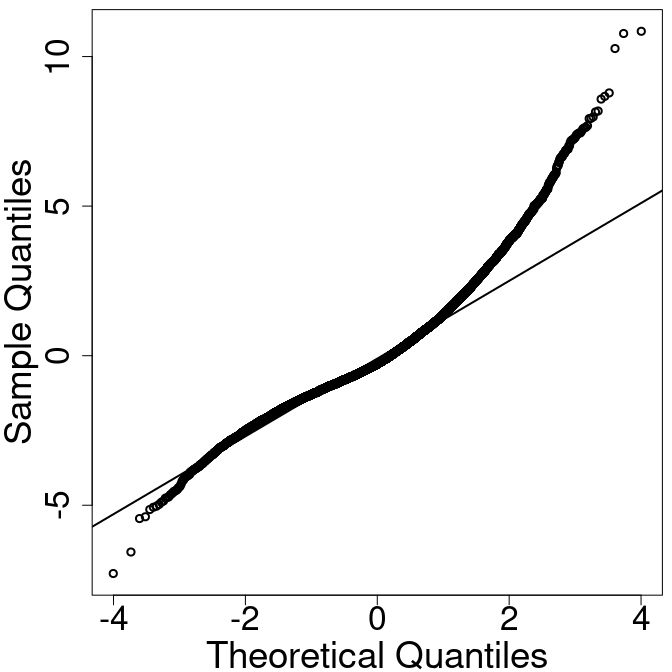}
\includegraphics[width=1\linewidth, height=6.25cm, keepaspectratio]{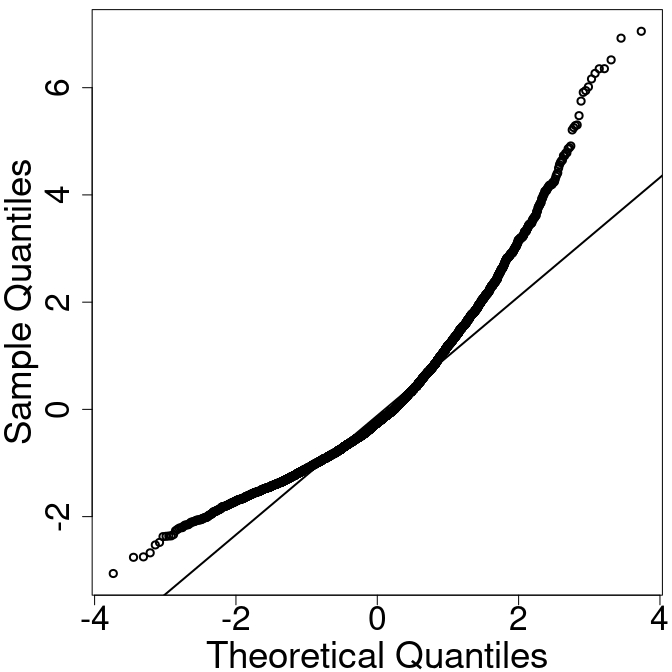}
\includegraphics[width=1\linewidth, height=6.25cm, keepaspectratio]{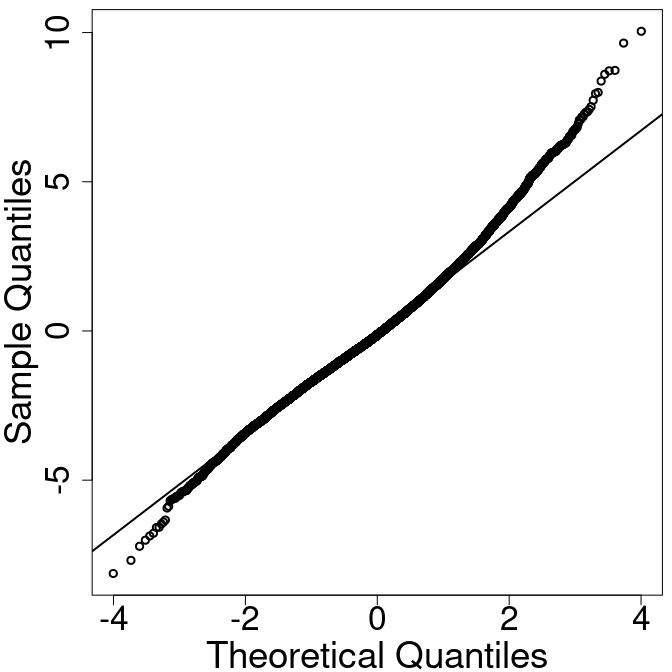}
\includegraphics[width=1\linewidth, height=6.25cm, keepaspectratio]{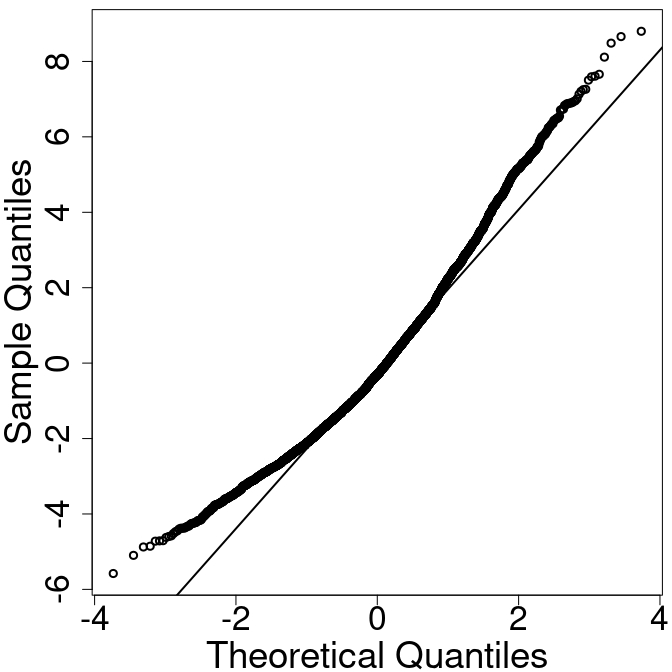}
\caption{Normal probability plots of level 1 (first column) and level 2 (second column) residuals from the LREM for SDQ internalizing (first row) and externalizing (second row) problems.}\label{fig:res}
\end{figure}

 \red{In order to highlight the practical relevance of the proposed methodology,} \red{we compare our findings with the parameter estimates of the \red{univariate} LREM and LQMM reported in Table \ref{tab:est_lqmm}. At first, we observe that the LQMM results are generally in line with our \red{findings}, except for $\tau = 0.25$. 
 Point estimates of the LREM and the LQMM at the median are not similar due to the asymmetry in SDQ distributions. This is consistent with the graphical analysis in Figure \ref{fig:data}, and highlights the importance of considering a quantile regression approach to assess the heterogeneous impact of risk factors across the distribution of children’s psychopathologies. To further show that modeling the conditional mean is an unreasonable approach, Figure \ref{fig:res} presents normal probability plots of level 1 and level 2 residuals of the fitted LREM models. These reveal the presence of potentially influential observations in the data, indicate severe departures from the Gaussian assumption of the random-intercepts model for both SDQ outcomes and show that residuals are skewed.}


 \red{In addition to that, both the LQMM and LREM} analyze children's disorders by fitting two univariate models separately and hence, \red{they disregard} the possible association between the SDQ scores. \textcolor{black}{In contrast}, one of the main benefits of the proposed multivariate approach is the possibility to study the \red{magnitude and direction of the} dependence structure between the responses at different \red{quantile levels of interest.} Following \cite{kotz2012laplace}, we can compute the correlation between \red{SDQ scores using \eqref{eq:corr} and understand whether their association structure becomes stronger for children with more pronounced problems.} 
 In particular, the estimated correlation coefficient, $r_{12}$, reported in Table \ref{tab:est} gives a measure of tail correlation and, consistently with the recent work of \cite{alfo2020m}, it indicates that internalizing and externalizing disorders are positively associated and this association increases with \red{the quantile level} $\tau$. \red{From the median to the 90-th percentile, the estimated coefficient rapidly increases from 0.211 to 0.738, which suggests that children \red{with} high levels of internalizing SDQ scores are more likely to experience or develop externalizing problems, and viceversa. Hence, children may present a constellation of symptoms comprised of both disorders which is aggravated in disadvantaged ones by the accumulation of risk factors.} This finding is also in line with the existence of positive covariation among psychiatric diagnoses (see \cite{lilienfeld2003comorbidity, liu2004childhood} and \cite{cicchetti2014developmental}). 

\begin{figure}[t]
\center
\includegraphics[width=1\linewidth, height=6.25cm, keepaspectratio]{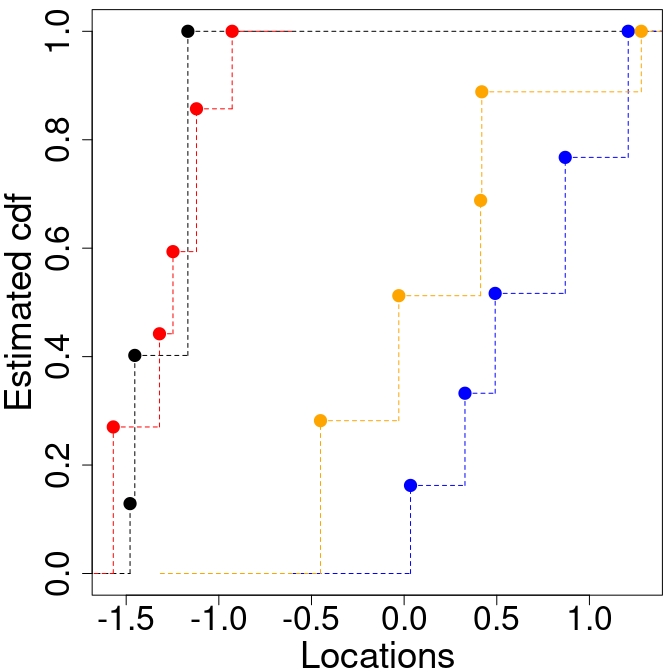}
\includegraphics[width=1\linewidth, height=6.25cm, keepaspectratio]{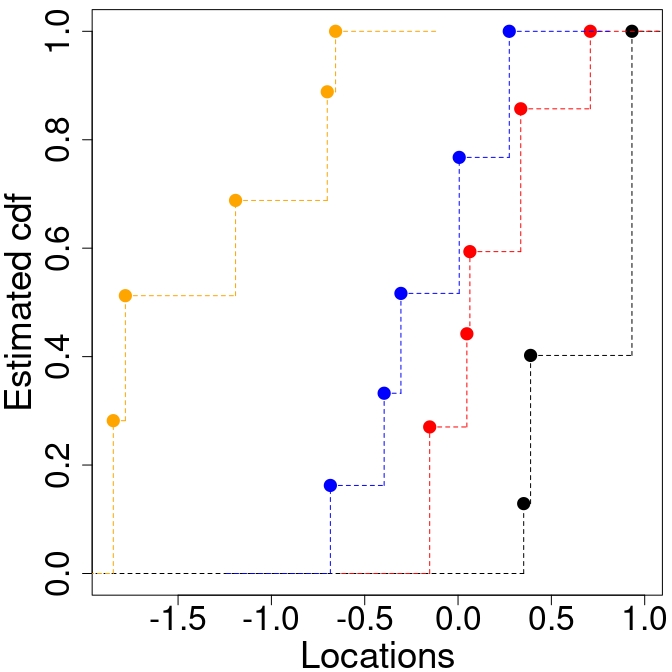}
\caption{Estimated cumulative density function of the discrete random slopes for SDQ internalizing (left) and externalizing (right) problems scores, at the 0.25 (black), 0.50 (red), 0.75 (blue) \red{and 0.90 (orange)} quantile levels.}\label{fig:ecdf}
\end{figure}

 \red{To further justify our empirical strategy, we conclude the analysis by reporting selected diagnostics for the fitted QMHMM. Firstly, } Figure \ref{fig:ecdf} shows the estimated marginal cumulative density functions of the \red{discrete} random slopes for both SDQ outcomes. In both plots, it is clear that the estimated distribution functions depart substantially from the Gaussian distribution, having pronounced asymmetries. \red{Hence, the underlying assumption of normally distributed random intercepts in the LQMM and LRE models is inappropriate. In contrast, the discrete mixture performs is more flexible and is able to accommodate possible departures from the Gaussianity assumption (\cite{alfo2017finite}). 
 \\
 Secondly, Tables \ref{tab:est} and \ref{tab:esthmm} summarize the estimated random intercepts $\boldsymbol{\alpha}$,} initial and transition probabilities of the hidden Markov chain. \red{Inference about the hidden Markov process gives additional insight into the \red{evolution of the SDQ scores over} time and the serial heterogeneity between subjects. In this application, the states are not only a tool for modeling time-dependence but also have a practical meaning. The transition matrices describe how, and how frequently, children move from low to high level of disorders and the random intercepts $\boldsymbol{\alpha}$ correspond to different severities of disorders in children}. 
 At first, it is worth noting that the estimated state-dependent intercepts $\boldsymbol{\alpha}$, 
 tend to increase when moving from lower to upper quantiles resulting in higher levels of children's disorders. For $\tau = (0.25, 0.25)$ (Panel A), the initial probability distribution defined on $\mathcal{S}$ is relatively uniformly distributed and the probability of not moving from states 1 and 4 is also very high, i.e. $\widehat{q}_{11} = 0.991$ and $\widehat{q}_{44} = 0.993$, respectively. This implies that almost every child in the lower tail of the outcomes distributions starts and maintains low-level disorders over time. If any transition is observed, units tend to move towards states 1 and 4 with lower intercepts and a reduction in juvenile developmental disorders with temporary jumps to moderate values of disorders. For $\tau = (0.50, 0.50)$ (Panel B), by looking at the initial probabilities one can see that half of the units ($\widehat{q}_{3} = 0.415$) start the study with low values of emotional and behavioral disorders and transitions between states are unlikely. When $\tau = (0.75, 0.75)$ (Panel C), the majority of children start with moderate values of developmental difficulties ($\widehat{q}_{1} + \widehat{q}_{2} > 0.80)$ and transitions to more severe disorders are more frequent. \red{Finally, by looking at the right tail of the SDQ scores distributions, $\tau = (0.90, 0.90)$ (Panel D), we conclude that, \red{among those children with high problems, states 2 and 3 are associated with even higher incidence of disorders.} 
 Around 60\% of children start the study \red{in state 1} and 91\% remain within the same class, with temporary changes \red{towards more severe level of emotional and behavioral disorders.}}
\begin{table}[!h]
\centering
 \smallskip 
 \resizebox{1.0\columnwidth}{!}{
\begin{tabular}{lcccccccc}
\toprule
$\bs \tau$-th quantile
& \multicolumn{2}{c}{(0.25, 0.25)}&\multicolumn{2}{c}{(0.50, 0.50)}&\multicolumn{2}{c}{(0.75, 0.75)}&\multicolumn{2}{c}{(0.90, 0.90)} \\
& \multicolumn{2}{c}{[G = 3, M = 5]}&\multicolumn{2}{c}{[G = 5, M = 4]}&\multicolumn{2}{c}{[G = 5, M = 3]}&\multicolumn{2}{c}{[G = 5, M = 3]} \\\cmidrule(r){2-3}\cmidrule(r){4-5}\cmidrule(r){6-7}\cmidrule(r){8-9}

Variable & SDQ$_{Int}$ & SDQ$_{Ext}$ & SDQ$_{Int}$ & SDQ$_{Ext}$ & SDQ$_{Int}$ & SDQ$_{Ext}$ & SDQ$_{Int}$ & SDQ$_{Ext}$ \\
\hline
Age year scal   & $\mathbf{-0.053} \; (0.010)$  & $\mathbf{-0.289} \; (0.016)$ & $\mathbf{-0.051} \; (0.009)$ & $\mathbf{-0.440} \; (0.011)$ & $-0.040 \; (0.021)$ & $\mathbf{-0.450} \; (0.031)$ & $-0.015 \; (0.017)$ & $\mathbf{-0.414} \; (0.029)$  \\
Age2 year scal  & $\mathbf{0.045} \; (0.005)$  & $\mathbf{0.193} \; (0.009)$  & $\mathbf{0.077} \; (0.011)$  & $\mathbf{0.208} \; (0.013)$  & $\mathbf{0.104} \; (0.018)$  & $\mathbf{0.287} \; (0.024)$ & $\mathbf{0.065} \; (0.014)$  & $\mathbf{0.233} \; (0.021)$  \\
ALE 11          & $\mathbf{0.022} \; (0.008)$  & $\mathbf{0.036} \; (0.016)$  & $\mathbf{0.086} \; (0.018)$  & $\mathbf{0.113} \; (0.019)$  & $\mathbf{0.116} \; (0.039)$  & $\mathbf{0.205} \; (0.055)$ & $\mathbf{0.202} \; (0.031)$  & $\mathbf{0.284} \; (0.048)$  \\
SED 4           & $\mathbf{0.070} \; (0.023)$  & $\mathbf{0.105} \; (0.045)$  & $\mathbf{0.175} \; (0.030)$  & $\mathbf{0.221} \; (0.030)$  & $\mathbf{0.201} \; (0.051)$  & $\mathbf{0.398} \; (0.076)$ & $\mathbf{0.217} \; (0.033)$  & $\mathbf{0.347} \; (0.059)$  \\
Kessm           & $\mathbf{0.090} \; (0.009)$  & $\mathbf{0.143} \; (0.012)$  & $\mathbf{0.167} \; (0.009)$  & $\mathbf{0.189} \; (0.012)$  & $\mathbf{0.208} \; (0.018)$  & $\mathbf{0.299} \; (0.025)$ & $\mathbf{0.257} \; (0.012)$  & $\mathbf{0.379} \; (0.020)$  \\
Degree          & $\mathbf{-0.350} \; (0.109)$ & $\mathbf{-0.894} \; (0.149)$ & $\mathbf{-0.526} \; (0.114)$ & $\mathbf{-1.482} \; (0.171)$ & $\mathbf{-0.703} \; (0.160)$ & $\mathbf{-1.267} \; (0.232)$ & $\mathbf{-0.856} \; (0.135)$ & $\mathbf{-1.530} \; (0.271)$ \\
GCSE            & $\mathbf{-0.217} \; (0.110)$ & $\mathbf{-0.582} \; (0.150)$ & $\mathbf{-0.352} \; (0.113)$ & $\mathbf{-0.430} \; (0.166)$ & $\mathbf{-0.413} \; (0.149)$ & $\mathbf{-0.427} \; (0.213)$ & $\mathbf{-0.524} \; (0.132)$ & $\mathbf{-0.671} \; (0.258)$ \\
White          & $-0.090 \; (0.061)$          & $-0.143 \; (0.097)$          & $-0.075 \; (0.149)$          & $0.059 \; (0.160)$           & $-0.216 \; (0.203)$          & $0.320 \; (0.303)$ & $-0.218 \; (0.127)$          & $0.125 \; (0.259)$           \\
Male            & $\mathbf{-0.062} \; (0.016)$ & $\mathbf{0.793} \; (0.030)$  & $0.027 \; (0.037)$           & $\mathbf{0.950} \; (0.045)$  & $0.082 \; (0.080)$           & $\mathbf{0.944} \; (0.119)$ & $\mathbf{0.258} \; (0.053)$  & $\mathbf{1.308} \; (0.094)$  \\
IMD        & $\mathbf{-0.022} \; (0.005)$ & $\mathbf{-0.036} \; (0.009)$ & $\mathbf{-0.025} \; (0.008)$ & $\mathbf{-0.027} \; (0.010)$ & $-0.027 \; (0.020)$          & $-0.045 \; (0.029)$ & $\mathbf{-0.054} \; (0.012)$ & $\mathbf{-0.079} \; (0.025)$ \\
Eth stratum & $-0.043 \; (0.159)$          & $-0.168 \; (0.174)$          & $0.122 \; (0.193)$           & $0.144 \; (0.220)$           & $0.174 \; (0.241)$           & $-0.025 \; (0.374)$ & $-0.063 \; (0.136)$          & $-0.345 \; (0.298)$          \\
Dis stratum & $-0.003 \; (0.031)$          & $0.003 \; (0.072)$           & $\mathbf{0.085} \; (0.042)$           & $\mathbf{0.106} \; (0.050)$  & $0.156 \; (0.111)$           & $\mathbf{0.337} \; (0.171)$ & $0.005 \; (0.064)$           & $\mathbf{0.203} \; (0.103)$           \\
\\
$\boldsymbol{\alpha}_1$ & $\mathbf{0.437} \; (0.001)$ & $\mathbf{1.040} \; (0.114)$ & $\mathbf{0.983} \; (0.013)$ & $\mathbf{4.769} \; (0.049)$ & $\mathbf{2.082} \; (0.016)$ & $\mathbf{2.954} \; (0.054)$ & $\mathbf{3.311} \; (0.005)$ & $\mathbf{5.078} \; (0.006)$ \\
$\boldsymbol{\alpha}_2$ & $\mathbf{3.930} \; (0.051)$  & $\mathbf{4.140} \; (0.089)$ & $\mathbf{2.480} \; (0.007)$  & $\mathbf{8.329} \; (0.077)$ & $\mathbf{2.735} \; (0.119)$ & $\mathbf{7.769} \; (0.014)$ & $\mathbf{9.104} \; (0.069)$ & $\mathbf{9.919} \; (0.103)$ \\
$\boldsymbol{\alpha}_3$ & $\mathbf{1.008} \; (0.005)$  & $\mathbf{6.362} \; (0.000)$ & $\mathbf{1.164} \; (0.016)$ & $\mathbf{1.954} \; (0.044)$& $\mathbf{6.957} \; (0.091)$  & $\mathbf{6.347} \; (0.201)$ & $\mathbf{3.848} \; (0.056)$ & $\mathbf{10.541} \; (0.212)$  \\ 
$\boldsymbol{\alpha}_4$ & $\mathbf{1.523} \; (0.130)$  & $\mathbf{1.281} \; (0.099)$ & $\mathbf{4.829} \; (0.008)$  & $\mathbf{4.670} \; (0.128)$  \\
$\boldsymbol{\alpha}_5$ & $\mathbf{0.477} \; (0.043)$ & $\mathbf{3.115} \; (0.013)$ & & \\

\\
$r_{12}$ & $\mathbf{0.356} \; (0.016)$ & & $\mathbf{0.211} \; (0.023)$ & & $\mathbf{0.602} \; (0.021)$ & & $\mathbf{0.738} \; (0.013)$ & \\ 
\\
$\ell ({\bf \Phi}_{\bs{\tau}})$ & -71665.9 & & -73173.9 & & -77223.6 & & -80517.3 & \\
$\#$ par & 69 & & 64 & & 55 & & 55 & \\ 
AIC & 143470.0 & & 146794.6 & & 154557.2 & & 161144.5 & \\
BIC & 143924.2 & & 146914.3 & & 154919.3 & & 161506.6 &\\

\bottomrule
\end{tabular}}
\caption{Point estimates with standard errors in parentheses for different quantile levels. Parameter estimates are displayed in boldface when significant at the standard 5\% level. 
}
\label{tab:est}
\end{table}

\newgeometry{margin=1.5cm}

\begin{sidewaystable}
\centering
 \smallskip 
 \resizebox{1.025\columnwidth}{!}{
\begin{tabular}{lcccccccccc}
\toprule
$\bs \tau$-th quantile & \multicolumn{2}{c}{(0.25, 0.25)}&\multicolumn{2}{c}{(0.50, 0.50)}&\multicolumn{2}{c}{LREM - Mean}&\multicolumn{2}{c}{(0.75, 0.75)}&\multicolumn{2}{c}{(0.90, 0.90)}
\\\cmidrule(r){2-3}\cmidrule(r){4-5}\cmidrule(r){6-7}\cmidrule(r){8-9}\cmidrule(r){10-11}
Variable & SDQ$_{Int}$ & SDQ$_{Ext}$ & SDQ$_{Int}$ & SDQ$_{Ext}$ & SDQ$_{Int}$ & SDQ$_{Ext}$ & SDQ$_{Int}$ & SDQ$_{Ext}$ & SDQ$_{Int}$ & SDQ$_{Ext}$\\
\hline
Intercept       & $\mathbf{1.808} \; (0.165)$  & $\mathbf{3.298} \; (0.233)$  & $\mathbf{2.313} \; (0.154)$  & $\mathbf{3.942} \; (0.241)$  & $\mathbf{2.575} \; (0.149)$  & $\mathbf{4.314} \; (0.221)$  & $\mathbf{2.900} \; (0.174)$  & $\mathbf{4.438} \; (0.234)$  & $\mathbf{3.123} \; (0.175)$  & $\mathbf{4.954} \; (0.248)$  \\
Age year scal   & $-0.000 \; (0.022)$          & $\mathbf{-0.391} \; (0.012)$ & $\mathbf{-0.037} \; (0.008)$ & $\mathbf{-0.472} \; (0.011)$ & $-0.002 \; (0.009)$          & $\mathbf{-0.450} \; (0.011)$ & $\mathbf{-0.036} \; (0.012)$ & $\mathbf{-0.488} \; (0.012)$ & $-0.016 \; (0.020)$          & $\mathbf{-0.525} \; (0.025)$ \\
Age2 year scal  & $0.000 \; (0.018)$           & $\mathbf{0.175} \; (0.009)$  & $\mathbf{0.053} \; (0.006)$  & $\mathbf{0.238} \; (0.006)$ & $\mathbf{0.083} \; (0.007)$  & $\mathbf{0.252} \; (0.009)$ & $\mathbf{0.103} \; (0.008)$  & $\mathbf{0.298} \; (0.009)$  & $\mathbf{0.111} \; (0.013)$  & $\mathbf{0.312} \; (0.015)$  \\
ALE 11          & $0.000 \; (0.018)$           & $\mathbf{0.120} \; (0.026)$  & $\mathbf{0.055} \; (0.015)$  & $\mathbf{0.072} \; (0.030)$ & $\mathbf{0.095} \; (0.016)$  & $\mathbf{0.114} \; (0.021)$ & $\mathbf{0.105} \; (0.021)$  & $\mathbf{0.113} \; (0.029)$  & $\mathbf{0.106} \; (0.049)$  & $\mathbf{0.185} \; (0.040)$  \\
SED 4           & $-0.000 \; (0.043)$          & $\mathbf{0.110} \; (0.042)$  & $\mathbf{0.181} \; (0.036)$  & $\mathbf{0.185} \; (0.054)$ & $\mathbf{0.138} \; (0.024)$  & $\mathbf{0.200} \; (0.032)$ & $\mathbf{0.231} \; (0.048)$  & $\mathbf{0.259} \; (0.038)$  & $\mathbf{0.204} \; (0.048)$  & $\mathbf{0.285} \; (0.093)$  \\
Kessm           & $0.000 \; (0.054)$           & $\mathbf{0.147} \; (0.014)$  & $\mathbf{0.146} \; (0.008)$  & $\mathbf{0.143} \; (0.026)$ & $\mathbf{0.161} \; (0.007)$  & $\mathbf{0.195} \; (0.009)$ & $\mathbf{0.187} \; (0.012)$  & $\mathbf{0.215} \; (0.016)$  & $\mathbf{0.230} \; (0.023)$  & $\mathbf{0.214} \; (0.028)$  \\
Degree          & $\mathbf{-1.020} \; (0.181)$ & $\mathbf{-1.407} \; (0.171)$ & $\mathbf{-0.564} \; (0.111)$ & $\mathbf{-1.383} \; (0.162)$ & $\mathbf{-0.687} \; (0.099)$ & $\mathbf{-1.583} \; (0.148)$ & $\mathbf{-0.483} \; (0.131)$ & $\mathbf{-1.246} \; (0.160)$ & $\mathbf{-0.390} \; (0.136)$ & $\mathbf{-1.182} \; (0.184)$ \\
GCSE            & $-0.020 \; (0.204)$          & $\mathbf{-0.640} \; (0.163)$ & $\mathbf{-0.392} \; (0.112)$ & $\mathbf{-0.391} \; (0.170)$ & $\mathbf{-0.372} \; (0.093)$ & $\mathbf{-0.680} \; (0.141)$ & $-0.117 \; (0.118)$          & $\mathbf{-0.317} \; (0.158)$ & $-0.043 \; (0.119)$          & $-0.210 \; (0.187)$          \\
White           & $\mathbf{-0.288} \; (0.121)$ & $0.009 \; (0.145)$           & $\mathbf{-0.275} \; (0.110)$ & $0.255 \; (0.157)$ & $\mathbf{-0.352} \; (0.091)$ & $0.121 \; (0.138)$          & $-0.132 \; (0.111)$          & $\mathbf{0.445} \; (0.146)$  & $-0.015 \; (0.126)$          & $\mathbf{0.713} \; (0.181)$  \\
Male            & $-0.000 \; (0.022)$          & $\mathbf{0.744} \; (0.093)$  & $0.061 \; (0.039)$           & $\mathbf{0.958} \; (0.073)$ & $\mathbf{0.127} \; (0.049)$  & $\mathbf{0.938} \; (0.075)$  & $\mathbf{0.232} \; (0.083)$  & $\mathbf{1.082} \; (0.087)$  & $\mathbf{0.319} \; (0.103)$  & $\mathbf{1.164} \; (0.118)$  \\
IMD        & $-0.000 \; (0.019)$          & $-0.027 \; (0.021)$          & $\mathbf{-0.033} \; (0.011)$ & $-0.009 \; (0.021)$ & $\mathbf{-0.044} \; (0.010)$ & $\mathbf{-0.048} \; (0.014)$         & $\mathbf{-0.048} \; (0.018)$ & $-0.006 \; (0.022)$          & $-0.026 \; (0.022)$          & $-0.054 \; (0.036)$          \\
Eth stratum & $-0.046 \; (0.097)$          & $-0.189 \; (0.196)$          & $0.153 \; (0.119)$           & $0.089 \; (0.191)$ & $0.169 \; (0.112)$           & $-0.016 \; (0.168)$          & $\mathbf{0.373} \; (0.139)$  & $0.216 \; (0.203)$           & $\mathbf{0.541} \; (0.141)$  & $0.281 \; (0.224)$           \\
Dis stratum & $-0.001 \; (0.040)$          & $0.058 \; (0.100)$           & $0.039 \; (0.056)$           & $0.095 \; (0.094)$ & $0.070 \; (0.061)$           & $\mathbf{0.226} \; (0.092)$          & $\mathbf{0.227} \; (0.076)$  & $\mathbf{0.379} \; (0.093)$  & $\mathbf{0.272} \; (0.093)$  & $\mathbf{0.535} \; (0.108)$  \\
$\sigma^2$ & 1.459 & 4.578 & 1.986 & 5.708 & 2.215 & 5.907 & 3.527 & 7.062 & 4.432 & 8.040 \\
\\
$\ell ({\bf \Phi}_{\bs{\tau}})$ & -32377.3 & -39048.1 & -34088.9 & -39125.8 & -34799.8 & -39229.7 & -36086.5 & -40174.2 & -37454.3 & -41192.0 \\ 
AIC & 64784.6 & 78126.3 & 68207.7 & 78281.6 & 69629.6 & 78489.9 & 72203.1 & 80378.4 & 74938.7 & 82414.0 \\ 
BIC & 64883.4 & 78225.0 & 68306.5 & 78380.3 & 69728.4 & 78588.2 & 72301.8 & 80477.2 & 75037.4 & 82512.7 \\ 
\bottomrule
\end{tabular}}
\caption{Univariate LREM for the mean and LQMM point estimates for internalizing and externalizing scores at the investigated quantile levels. Standard errors are in parentheses. Parameter estimates are displayed in boldface when significant at the standard 5\% level.}
\label{tab:est_lqmm}
\end{sidewaystable}
\restoregeometry

\begin{table}[h]
\centering
 \smallskip 
 \resizebox{1.0\columnwidth}{!}{
 \setlength\tabcolsep{13pt}
\begin{tabular}{lcccccccccc}
\toprule
& 1 & 2 & 3 & 4 & 5 \\
\toprule
Panel A: $\tau = (0.25, 0.25)$ & \\
\\
$\bs q$ & $0.201 \; (0.017)$ & $0.111 \; (0.013)$ & $0.269 \; (0.017)$ & $0.165 \; (0.017)$ & $0.254 \; (0.020)$ \\
1 & $0.991 \; (0.016)$ & $0.000 \; (0.000)$ & $0.000 \; (0.000)$ & $0.009 \; (0.004)$ & $0.000 \; (0.029)$ \\
2 & $0.000 \; (0.002)$ & $0.883 \; (0.044)$ & $0.007 \; (0.019)$ & $0.110 \; (0.008)$ & $0.000 \; (0.012)$ \\
3 & $0.000 \; (0.000)$ & $0.089 \; (0.012)$ & $0.723 \; (0.032)$ & $0.000 \; (0.000)$ & $0.187 \; (0.014)$ \\
4 & $0.003 \; (0.015)$ & $0.004 \; (0.042)$ & $0.000 \; (0.001)$ & $0.993 \; (0.009)$ & $0.000 \; (0.013)$ \\
5 & $0.096 \; (0.003)$ & $0.016 \; (0.009)$ & $0.027 \; (0.029)$ & $0.029 \; (0.000)$ & $0.832 \; (0.034)$ \\
\\
Panel B: $\tau = (0.50, 0.50)$ \\
\\
$\bs q$ & $0.287 \; (0.019)$ & $0.160 \; (0.016)$ & $0.415 \; (0.018)$ & $0.138 \; (0.014)$ \\
1 & $0.863 \; (0.036)$ & $0.034 \; (0.034)$ & $0.071 \; (0.007)$ & $0.032 \; (0.024)$ \\
2 & $0.053 \; (0.017)$ & $0.850 \; (0.038)$ & $0.000 \; (0.000)$ & $0.096 \; (0.013)$ \\
3 & $0.006 \; (0.031)$ & $0.000 \; (0.000)$ & $0.967 \; (0.013)$ & $0.027 \; (0.034)$ \\
4 & $0.049 \; (0.015)$ & $0.015 \; (0.024)$ & $0.067 \; (0.011)$ & $0.868 \; (0.036)$ \\
\\
Panel C: $\tau = (0.75, 0.75)$ \\
\\
$\bs q$ & $0.541 \; (0.021)$ & $0.293 \; (0.017)$ & $0.165 \; (0.015)$ \\
1 & $0.942 \; (0.011)$ & $0.006 \; (0.020)$ & $0.052 \; (0.029)$ \\
2 & $0.072 \; (0.004)$ & $0.825 \; (0.024)$ & $0.103 \; (0.020)$ \\
3 & $0.185 \; (0.010)$ & $0.109 \; (0.015)$ & $0.707 \; (0.031)$ \\
\\
\red{Panel D: $\tau = (0.90, 0.90)$} \\
$\bs q$ & $0.606 \; (0.015)$ & $0.152 \; (0.011)$ & $0.242 \; (0.014)$ \\
1 & $0.913 \; (0.008)$ & $0.067 \; (0.025)$ & $0.019 \; (0.019)$ \\
2 & $0.292 \; (0.007)$ & $0.605 \; (0.026)$ & $0.103 \; (0.012)$ \\
3 & $0.147 \; (0.004)$ & $0.099 \; (0.015)$ & $0.754 \; (0.022)$ \\
\bottomrule
\end{tabular}}
\caption{Initial probabilities, $\bs q$, and transition probabilities, $\bs Q$, estimates for different quantiles.}
\label{tab:esthmm}
\end{table}

\section{Conclusions}\label{sec:con}
Longitudinal data allows us to understand the evolution of a certain phenomenon \red{over} time. In this context, it becomes of crucial importance to determine an appropriate modeling framework to assess the effects of unobserved factors and hidden heterogeneity which can be either time-invariant or time-varying; ignoring these factors may induce bias and lead to \red{invalid} conclusions. Moreover, the literature on this topic which is traditionally focused on the conditional mean, might not provide a good summary of the \red{response} distribution. 
 To account for \red{the} complex data structure, this work generalizes the \red{multivariate} quantile approach of \cite{petrella2019joint} \red{for the analysis of} multivariate longitudinal data by combining the features of quantile regression and MHMMs (\cite{altman2007mixed}). \red{The proposed model allows for the quantile-specific effects to be quantified and jointly modeling of several outcomes. The model further allows for} 
 different sources of heterogeneity \red{to} be distinguished, i.e. between individual heterogeneity and time heterogeneity are modeled through the state-specific effects.
 In order to avoid possibly misleading inferences \red{caused by erroneous} assumption on the random effects distribution, we rely on the NPML estimation theory and we approximate \red{this distribution by} a multivariate discrete latent variable. 

As illustrated in the real data application, the proposed method models simultaneously the quantiles of children's emotional and behavioral disorders as a function of demographic and socio-economics risk factors. The results show that behavioral and emotional difficulties are mainly affected by the family poverty conditions and mother's characteristics. Such effects are much stronger in the upper tail of the response distribution, i.e. for those children experiencing more severe internalizing and externalizing problems. In addition, the analysis reveals moderate levels of \red{codependency} between internalizing and externalizing disorders, \red{that cannot be detected by univariate models}. 

The methodology can be further extended to \red{allow} for a non-homogeneous hidden Markov process where transition probabilities are allowed to depend on covariates. 
Finally, the hidden Markov chain implicitly assumes that the sojourn time, i.e. the number of consecutive time points that the process spends in a given state, is geometrically distributed. As a further generalization of this work one may consider a semi-Markov process which is designed to relax this condition by allowing the sojourn time to be modeled by more flexible 
distributions.

\cleardoublepage
\bibliographystyle{agsm}
\newpage
{\bibliography{biblio}}

\clearpage
\section*{Appendix A}\label{appendixA}

\noindent \textbf{Proof of Proposition \ref{prop:completeloglik}}

Under the constraints imposed on $\bs {\tilde\xi}$ and $\bs \Lambda$, the representation in \eqref{eq:mixtureALD} implies that:
\begin{equation}\label{eq:hier1}
\bs Y \mid \tilde C = \tilde c \sim {\cal N}_p(\boldsymbol{\mu} + \bs D\tilde{\bs \xi} \tilde c \, , \, \tilde c \bs D { \bs \Sigma} \bs D), \qquad \tilde C \sim \textnormal{Exp} (1).
\end{equation}
This implies that the joint density function of $\bs Y$ and $\tilde C$ is:
\begin{equation}\label{eq:joint}
f_{\bs Y,\tilde C}(\bs y,\tilde c)= \frac{\exp{\left\{ (\bs y - \boldsymbol{\mu}) ' \bs D^{-1}\bs \Sigma^{-1} \tilde{\bs \xi} \right\}}}{(2 \pi)^{p/2} \mid \bs D { \bs \Sigma} \bs D \mid^{1/2}} \left(  \tilde c^{-p/2} \exp{\left\{ -\frac{1}{2}   \frac{\tilde m}{\tilde c} - \frac{1}{2} \tilde c (\tilde d +2)    \right\}}\right).
\end{equation}
Then, the complete log-likelihood function (up to additive constant terms) can be written as follows:
\begin{equation}\label{eq:completeLik}
\begin{split}
\log \ell_c (\boldsymbol{\Phi}_\tau \mid \bs y, \bs x, \tilde{\bs c}, \bs s, \bs b) & = \sum_{i=1}^N \Bigg\{ \sum_{g=1}^G w_{ig} \log \pi_g + \sum_{j=1}^M u_{i1j} \log q_j + \sum_{t=2}^{T_i} \sum_{j=1}^M \sum_{k=1}^M v_{itjk} \log q_{jk} \\
& + \sum_{t=1}^{T_i} \sum_{j=1}^M \sum_{g=1}^G z_{itjg} \log f_{\bs{Y}, \tilde C} (\bs y_{it}, \tilde c_{it} \mid \bs x_{it}, S_{it} = j, \bs{b}_g) \Bigg\}.
\end{split}
\end{equation}
By substituting \eqref{eq:joint} in \eqref{eq:completeLik}, we obtain:
\begin{equation}\label{eq:completeexpLik}
\begin{split}
\ell_c({\bf \Phi}_{\bs{\tau}}) & = \sum_{i=1}^N \Bigg\{ \sum_{g=1}^G w_{ig} \log \pi_g + \sum_{j=1}^M u_{i1j} \log q_j + \sum_{t=2}^{T_i} \sum_{j=1}^M \sum_{k=1}^M v_{itjk} \log q_{jk} \\
& - \frac{1}{2} T_i \log \mid \bs D {\bs \Sigma} \bs D \mid + \sum_{t=1}^{T_i} \sum_{j=1}^M \sum_{g=1}^G z_{itjg} (\bs Y_{it} - \boldsymbol{\mu}_{it})' \bs D^{-1} {\bs \Sigma}^{-1}\tilde {\bs \xi} \\
& - \frac{1}{2} \sum_{t=1}^{T_i} \sum_{j=1}^M \sum_{g=1}^G z_{itjg} \frac{1}{\tilde C_{itjg}} (\bs Y_{it} - \boldsymbol{\mu}_{it})' (\bs D {\bs \Sigma} \bs D)^{-1} (\bs Y_{it} - \boldsymbol{\mu}_{it}) \\
& - \frac{1}{2} \tilde{{\bs \xi}}' {\bs \Sigma^{-1}} \tilde {\bs \xi} \sum_{t=1}^{T_i} \sum_{j=1}^M \sum_{g=1}^G z_{itjg} \tilde C_{itjg} \Bigg\}.
\end{split}
\end{equation}

\medskip

\noindent \textbf{Proof of Proposition \ref{prop:expectedloglik}}

The E-step of the EM algorithm considers the conditional expectation of the complete log-likelihood function given the observed data and the current parameter estimates $\hat{{\bf \Phi}}^{(r-1)}_\tau$. The conditional expectations of $w_{ig}, u_{itj}, v_{itjk}$ and $z_{itjg}$ can be computed using standard arguments in the HMM literature as shown in \eqref{eq:posterior}. To compute the conditional expectation of $\tilde C$ and $\tilde C^{-1}$, in the E-step of the EM algorithm, $\tilde C$ is treated as an additional latent variable and, hence, not observable. 
Using the joint distribution of $\bs Y$ and $\tilde C$ derived in \eqref{eq:joint} and the MAL density of $\bs Y$ given in \eqref{eq:MALdensity}, we have that:
\begin{equation}
f_{\tilde C}(\tilde C \mid \bs Y = \bs y) = \frac{f_{\tilde C,\bs Y}(\tilde c,\bs y)}{f_{\bs Y}(\bs y)} = \frac{\tilde c^{-p/2} \left(\frac{2+\tilde d}{\tilde m} \right)^{\nu/2} \exp{\left\{  -\frac{\tilde m}{2 \tilde c}- \frac{\tilde c (2+\tilde d)}{2}   \right\}}}{2 K_{\nu}\left( \sqrt{(2+\tilde d)\tilde m} \right)},
\end{equation}
which corresponds to a Generalized Inverse Gaussian (GIG) distribution with parameters $\nu, {2+\tilde d}, \tilde{m_i}$, i.e.\footnote{The pdf of a GIG($p,a,b$) distribution is defined as $f_{GIG}(x; p,a,b)= \frac{\left(\frac{a}{b}\right)^{p/2}}{2K_{p}(\sqrt{ab})} x^{p-1} e^{-\frac{1}{2}\left( ax +bx^{-1}  \right)}$, with $a>0$, $b>0$ and $p \in {\cal R}$.}
\begin{equation}
f_{\tilde C}(\tilde C \mid \bs Y = \bs y) \sim \mbox{GIG}\left(\nu, \tilde d+2, \tilde{m}\right).
\end{equation}
It follows that 
\begin{equation}\label{eq:w1}
\mathbb{E}[\tilde C \mid \cdot] = \left(  \frac{\hat {\tilde {m}}}{2+\hat {\tilde d}}  \right)^{\frac{1}{2}} \frac{K_{\nu +1}\left( \sqrt{(2+\hat{\tilde d})\hat {\tilde {m}}} \right)}{K_{\nu}\left(   \sqrt{(2+\hat {\tilde d}) \hat {\tilde {m}}}\right)}
\end{equation}
and 
\begin{equation}\label{eq:w2}
\mathbb{E}[\tilde C^{-1} \mid \cdot] = \left( \frac{2+\hat {\tilde d}}{\hat {\tilde {m}}} \right)^{\frac{1}{2}}  \frac{K_{\nu +1}  \left( \sqrt{(2+\hat {\tilde d})\hat {\tilde {m}}} \right)}{K_{\nu}  \left( \sqrt{(2+\hat {\tilde d})\hat {\tilde {m}}} \right)} - \frac{2 \nu}{\hat {\tilde {m}}}.
\end{equation}
Denoting the two conditional expectations in \eqref{eq:w1} and \eqref{eq:w2} by $\hat{\tilde c}$ and $\hat{\tilde z}$ respectively, concludes the proof.

%


\medskip

\noindent \textbf{Proof of Proposition \ref{prop:Mupdates}}

Imposing the first order conditions on \eqref{eq:OF2} with respect to each component of the set ${\bf \Phi}_{\bs{\tau}}$, gives the parameter estimates in \eqref{eq:Mupdates}, \eqref{eq:focBeta} and \eqref{eq:focSigma}. However, there is not closed formula solution to update the elements of the scale matrix $\bs D$; hence, the M-step update requires using numerical optimization techniques to maximize \eqref{eq:OF2}. A considerable disadvantage of this procedure is the necessary high computational effort which could be very time-consuming. For this reason, we utilize a simpler estimator for the scale parameters $d_j, j = 1,\dots,p$ which follows directly from the fact that all marginals of the MAL distribution are univariate AL distributions (see \cite{yu2005three} and \cite{marino2018mixed}): 
\begin{equation}\label{eq:focDapp}
\hat{d}_j = \frac{1}{\sum_{i=1}^N T_i} \sum_{i=1}^N \sum_{t=1}^{T_i} \sum_{g=1}^G \sum_{k=1}^M \hat{z}_{itkg} \rho_\tau (Y_{it}^{(j)} - \hat{{\mu}}^{(j)}_{it}).
\end{equation}

\clearpage
\section*{\red{Appendix B}}\label{appendixB}
In this Appendix we conduct a simulation study to evaluate the finite sample properties of the proposed method and show that the introduced methodology represents a valid procedure to estimate the quantile regression coefficients. This simulation exercise addresses the following questions. First, we consider different distributional choices for the error term to study the performance of the model in the presence of non-Gaussian errors. Second, we evaluate the robustness of the non-parametric approach to non-Gaussian distributions for the subject-specific, random coefficients. \red{Finally, we analyze the performance of penalized likelihood criteria in selecting the optimal number of mixture components $G$ and hidden states $M$.}


We consider two sample sizes $N=(100, 200)$ and two longitudinal lengths $T_i = T = (5, 10)$, for all $i = 1,\dots,N$, for a continuous response variable of dimension $p=2$ and two explanatory variables $X^{(1)}_{it} \sim \mathcal{N}(0,1)$ and $X^{(2)}_{it} \sim \textnormal{Ber}(0.5)$. The observations are generated from a two state homogeneous Markov chain, i.e. $M = 2$, using the following data generating process:
\begin{equation}\label{eq:mhmms}
\mathbf{Y}_{it} = \mathbf{X}_{it} \boldsymbol{\beta} + \mathbf{Z}_{it} \mathbf{b}_i + \mathbf{W}_{it} \boldsymbol{\alpha}_{S_{it}} + \epsilon_{it}.
\end{equation} 
Regarding the hidden Markov chain, the simulation scheme is similar to the one adopted by \cite{marino2018mixed}. The true values of the fixed, $\boldsymbol{\beta}$, state dependent parameters, $\boldsymbol{\alpha} = (\boldsymbol{\alpha}_1, \dots, \boldsymbol{\alpha}_M)$ and the initial probabilities, $\bs{q}$, and transition probabilities, $\bs{Q}$, are given by, respectively: 
\begin{equation}\label{betatrue}
\boldsymbol{\beta} = 
\bqmatrix  2 & -0.8\\
            -1.4 & 3.0\eqmatrix, \quad
            \boldsymbol{\alpha}= 
\bqmatrix   5 & -2\\
            -5 & 2\eqmatrix, \quad
            \bs{Q} = 
            \bqmatrix   0.8 & 0.2\\
            0.2 & 0.8\eqmatrix, \quad
            \quad
            \bs{q} = 
            \bqmatrix  0.7 & 0.3\eqmatrix.  
\end{equation}
We consider a time-varying random intercept by setting $\mathbf{W}_{it} = \mathbf{1}$ and a random slope $\mathbf{Z}_{it} = X^{(1)}_{it}$. Hence, $\mathbf{b}_i$ are time-constant random slopes that capture individual departures from the marginal effect $\boldsymbol{\beta}$. \textcolor{black}{For each sample size, two different simulation scenarios for the error distributions and for the random coefficients distributions are considered:
\begin{enumerate}[label=]
\item ($\mathcal{N} - \mathcal{N}$): $\mathbf{b}_i$ represent i.i.d. draws from a standard bivariate Gaussian with variance-covariance matrix, $\bs \Omega = \bigl( \begin{smallmatrix}1 & 0.25\\ 0.25 & 1\end{smallmatrix}\bigr)$ and the error terms, $\epsilon_{{it}}$, are generated from a bivariate Normal random variable with zero mean vector and variance-covariance matrix equal to $\bs{\tilde \Omega}$;\medskip
\item ($\mathcal{T} - \mathcal{T}$): $\mathbf{b}_i$ are sampled from a bivariate Student t with 3 degrees of freedom, centered around zero and scale matrix $\bs \Omega = \bigl( \begin{smallmatrix}1 & 0.25\\ 0.25 & 1\end{smallmatrix}\bigr)$ while, $\epsilon_{{it}}$ are generated from a bivariate Student t distribution with 3 degrees of freedom, zero mean and scale matrix $\bs{\tilde \Omega}$.
\end{enumerate}
Each simulation scenario is repeated twice by generating the errors $\epsilon_{{it}}$ with low ($\bs{\tilde \Omega} = \bigl( \begin{smallmatrix}1 & 0.3\\ 0.3 & 1\end{smallmatrix}\bigr)$) and high ($\bs{\tilde \Omega} = \bigl( \begin{smallmatrix}1 & 0.8\\ 0.8 & 1\end{smallmatrix}\bigr)$) correlation between the responses of unit $i$ at a given time $t$.}

To fit the proposed model, we consider a varying number of mixture components $G = (2, \dots, 10)$ and retained the model with the lowest BIC value. We analyze three different quantile levels: in the first case, we assume $\bs \tau=(0.50.0.50)$; in the second one, we set $\bs{\tau}=(0.25, 0.25)$ and in the third, we set $\bs{\tau}=(0.75, 0.75)$. For each model, we carry out $B=250$ Monte Carlo replications and report the following indicators. The Average Relative Bias (ARB) defined as:
\begin{equation}
ARB (\hat{{\theta}}_\tau) = \frac{1}{B} \sum_{b=1}^B \frac{(\hat{{\theta}}^{(b)}_\tau - {{\theta}}_\tau)}{{{\theta}}_\tau} \times 100,
\end{equation}
where $\hat{{\theta}}^{(b)}_\tau$ is the estimated parameter at quantile level $\tau$ for the $b$-th replication and ${{\theta}}_\tau$ is the corresponding ``true'' value of the parameter.
 Secondly, the Root Mean Square Error (RMSE) of model parameters averaged across the $B$ simulations:
\begin{equation}
RMSE (\hat{{\theta}}_\tau) = \sqrt{\frac{1}{B} \sum_{b=1}^B (\hat{{\theta}}^{(b)}_\tau - {{\theta}}_\tau)^2 }.
\end{equation} 
Tables \ref{tab:s1} and \ref{tab:s2} report the results for the fixed parameters $\boldsymbol{\beta}$ and state-specific coefficients $\boldsymbol{\alpha}$.
 
 As can be \textcolor{black}{noted}, the proposed model under the Normal and the Student t error distributions \textcolor{black}{is} able to recover the true fixed parameters and state-dependent intercept values for both low (Panels A) and high (Panels B) degree of dependence. Not surprisingly, the bias effect is quite small when we analyze the median levels (see columns 1 and 4). As the quantile levels become more extreme (see columns 2, 3, 5 and 6), the ARB slightly increases but it still remains reasonably small. Such a differences are due to the reduced amount of information in the tails of the distribution. However, both the ARB and the RMSE tend to decrease with increasing sample sizes and number of measurement occasions. Also, \textcolor{black}{under the scenario where} $\bs{b}_i \sim \mathcal{T}_2 (\bs{0}, \bs \Omega)$ and $\epsilon_{{it}} \sim \mathcal{T}_2 (\bs{0}, \bs{\tilde \Omega})$, the heavier tails of the Student t contribute to higher ARB and RMSE especially at the 25-th and 75-th percentiles. Concerning the hidden process, it is worth noting that we observe sensible differences in terms of efficiency for the state dependent parameters $\boldsymbol{\alpha}$. Given the true values of $\bs Q$ and $\bs q$, most of the units are in the first state of the latent Markov chain, sharing the common intercept value $\boldsymbol{\alpha}_1$. Hence, the intercept corresponding to the second state $\boldsymbol{\alpha}_2$ is estimated with lower precision due to lack of transitions from one state to the other. However, when the number of repeated measurements increases, we observe more frequent transitions towards the second state with the effect of reducing the RMSE. Again, such difference is more evident in the tails of the distribution. These findings are generally consistent with the ones in \cite{marino2018mixed}.

 \red{
 To evaluate the performance of the model selection procedure described in Section \ref{sec:est},
 we considered the same simulation experiment with $N=200$, $T=10$, $M=2$, $\tau = (0.50, 0.50)$ and $B = 100$. Following \cite{marino2018mixed}, for each of the simulated dataset we fit the QMHMM for $G = (2, \dots, 8)$ and $M = (2, \dots, 4)$, and select the optimal value of the pair $(G,M)$ by using the AIC (\cite{akaike1998information}) and BIC in \eqref{eq:BIC}. Because the time-constant random slopes $\mathbf{b}_i$ in \eqref{eq:mhmms} are generated from continuous distributions, we only report the distribution of absolute frequencies of the hidden states $M$ selected by the two penalized likelihood criteria. 
 Table \ref{tab:s3} summarizes the results. 
 \\
 As one can see, the BIC works well and outperform the AIC, with an average of correctly identified number of hidden states of more than the 80\% across all simulation scenarios and levels of correlation. Furthermore, regardless of the distributional assumptions on the random slopes or on the error terms, the BIC captures the serial heterogeneity in the data in a more parsimonious manner compared to the AIC, \red{hence offering easier interpretation about} unobserved heterogeneity.}


\begin{table}[htbp]
 \center
 \smallskip 
 \resizebox{1.0\columnwidth}{!}{%
 \setlength\tabcolsep{15pt}
 
\begin{tabular}{lcccccc}
\toprule
& \multicolumn{3}{c}{($\mathcal{N} - \mathcal{N}$)} & \multicolumn{3}{c}{($\mathcal{T} - \mathcal{T}$)}\\\cmidrule(r){2-4}\cmidrule(r){5-7}
$\bs \tau $ & $(0.50,0.50)$ & $(0.25,0.25)$ & $(0.75,0.75)$ & $(0.50,0.50)$ & $(0.25,0.25)$ & $(0.75,0.75)$ \\
\hline
 Panel A: $\rho_{12} = 0.3$ & \\
\\
$\beta_{11}$  & $0.386 \; (0.114)$  & $0.316 \; (0.110)$  & $0.470 \; (0.116)$  & $-0.310 \; (0.168)$ & $-0.325 \; (0.175)$ & $-0.214 \; (0.174)$ \\
$\beta_{12}$  & $0.716 \; (0.122)$  & $1.317 \; (0.123)$  & $0.912 \; (0.125)$  & $0.951 \; (0.158)$  & $0.313 \; (0.166)$  & $0.871 \; (0.173)$  \\
$\beta_{21}$  & $-0.281 \; (0.071)$ & $0.942 \; (0.081)$  & $-2.603 \; (0.091)$ & $0.546 \; (0.075)$  & $2.137 \; (0.102)$  & $-1.752 \; (0.100)$ \\
$\beta_{22}$  & $-0.019 \; (0.073)$ & $-0.652 \; (0.091)$ & $0.838 \; (0.089)$  & $0.044 \; (0.083)$  & $-1.354 \; (0.111)$ & $1.163 \; (0.103)$  \\
$\alpha_{11}$ & $-0.028 \; (0.057)$ & $-0.252 \; (0.063)$ & $0.017 \; (0.070)$  & $0.033 \; (0.058)$  & $-0.201 \; (0.078)$ & $0.535 \; (0.091)$ \\
$\alpha_{12}$ & $0.397 \; (0.052)$  & $0.387 \; (0.069)$  & $0.042 \; (0.062)$  & $-0.180 \; (0.068)$ & $1.215 \; (0.082)$  & $-1.459 \; (0.093)$ \\
$\alpha_{21}$ & $-0.165 \; (0.074)$ & $-0.034 \; (0.082)$ & $-0.096 \; (0.080)$ & $0.179 \; (0.081)$  & $0.763 \; (0.113)$  & $0.066 \; (0.097)$  \\
$\alpha_{22}$ & $-0.120 \; (0.075)$ & $-0.057 \; (0.083)$ & $0.293 \; (0.084)$  & $0.011 \; (0.078)$  & $-1.862 \; (0.108)$ & $0.795 \; (0.095)$  \\

\\
 Panel B: $\rho_{12} = 0.8$ & \\
\\
$\beta_{11}$  & $0.281 \; (0.113)$  & $0.178 \; (0.111)$  & $0.294 \; (0.116)$  & $-0.339 \; (0.179)$ & $-0.668 \; (0.185)$ & $-0.655 \; (0.188)$ \\
$\beta_{12}$  & $0.951 \; (0.126)$  & $1.170 \; (0.127)$  & $0.700 \; (0.128)$  & $1.413 \; (0.165)$  & $1.876 \; (0.180)$  & $1.946 \; (0.173)$  \\
$\beta_{21}$  & $-0.522 \; (0.073)$ & $2.041 \; (0.087)$  & $-3.086 \; (0.091)$ & $0.282 \; (0.075)$  & $3.756 \; (0.108)$  & $-3.297 \; (0.113)$ \\
$\beta_{22}$  & $0.175 \; (0.074)$  & $-1.205 \; (0.094)$ & $1.502 \; (0.094)$  & $-0.106 \; (0.081)$ & $-2.019 \; (0.119)$ & $1.717 \; (0.119)$  \\
$\alpha_{11}$ & $-0.054 \; (0.055)$ & $-0.438 \; (0.073)$ & $0.168 \; (0.068)$  & $0.012 \; (0.063)$  & $-0.695 \; (0.094)$ & $0.813 \; (0.097)$ \\
$\alpha_{12}$ & $0.302 \; (0.052)$  & $1.255 \; (0.078)$  & $-0.434 \; (0.067)$ & $0.050 \; (0.068)$  & $1.997 \; (0.093)$  & $-2.190 \; (0.099)$ \\
$\alpha_{21}$ & $-0.031 \; (0.071)$ & $0.170 \; (0.083)$  & $-0.383 \; (0.077)$ & $0.028 \; (0.086)$  & $1.068 \; (0.121)$  & $-0.582 \; (0.118)$ \\
$\alpha_{22}$ & $-0.211 \; (0.076)$ & $-0.457 \; (0.081)$ & $1.068 \; (0.081)$  & $0.012 \; (0.082)$  & $-2.349 \; (0.115)$ & $1.873 \; (0.112)$  \\

 \bottomrule
\end{tabular}}
\caption{ARB and RMSE (in brackets) for longitudinal and state-parameter estimates with a sample size $N = 100$ and length of longitudinal sequences $T = 5$.}
\label{tab:s1}
\end{table}

\begin{table}[htbp]
 \center
 \smallskip 
 \resizebox{1.0\columnwidth}{!}{%
 \setlength\tabcolsep{15pt}
 
\begin{tabular}{lcccccc}
\toprule
& \multicolumn{3}{c}{($\mathcal{N} - \mathcal{N}$)} & \multicolumn{3}{c}{($\mathcal{T} - \mathcal{T}$)}\\\cmidrule(r){2-4}\cmidrule(r){5-7}
$\bs \tau$ & $(0.50,0.50)$ & $(0.25,0.25)$ & $(0.75,0.75)$ & $(0.50,0.50)$ & $(0.25,0.25)$ & $(0.75,0.75)$ \\
\hline
 Panel A: $\rho_{12} = 0.3$ & \\
\\
$\beta_{11}$  & $0.173 \; (0.077)$  & $0.008 \; (0.075)$  & $-0.004 \; (0.075)$ & $0.098 \; (0.118)$  & $-0.076 \; (0.120)$ & $-0.010 \; (0.120)$ \\
$\beta_{12}$  & $-0.432 \; (0.076)$ & $-0.273 \; (0.076)$ & $0.031 \; (0.076)$  & $-1.486 \; (0.107)$ & $-0.503 \; (0.109)$ & $-1.429 \; (0.106)$ \\
$\beta_{21}$  & $-0.291 \; (0.042)$ & $1.947 \; (0.056)$  & $-2.187 \; (0.055)$ & $-0.108 \; (0.044)$ & $1.628 \; (0.060)$  & $-2.230 \; (0.066)$ \\
$\beta_{22}$  & $0.066 \; (0.036)$  & $-1.013 \; (0.056)$ & $1.052 \; (0.055)$  & $-0.098 \; (0.043)$ & $-1.165 \; (0.069)$ & $1.280 \; (0.070)$  \\
$\alpha_{11}$ & $-0.096 \; (0.032)$ & $-0.197 \; (0.041)$ & $0.241 \; (0.041)$  & $0.030 \; (0.035)$  & $-0.278 \; (0.047)$ & $0.551 \; (0.057)$  \\
$\alpha_{12}$ & $0.116 \; (0.032)$  & $0.409 \; (0.041)$  & $-0.657 \; (0.039)$ & $0.062 \; (0.036)$  & $1.425 \; (0.059)$  & $-1.398 \; (0.063)$ \\
$\alpha_{21}$ & $-0.038 \; (0.037)$ & $0.262 \; (0.052)$  & $-0.296 \; (0.047)$ & $0.061 \; (0.039)$  & $0.666 \; (0.067)$  & $-0.349 \; (0.058)$ \\
$\alpha_{22}$ & $-0.040 \; (0.038)$ & $-0.862 \; (0.047)$ & $0.372 \; (0.042)$  & $-0.021 \; (0.038)$ & $-1.695 \; (0.067)$ & $1.568 \; (0.070)$  \\

 \\
 Panel B: $\rho_{12} = 0.8$ & \\
\\
$\beta_{11}$  & $0.040 \; (0.075)$  & $-0.028 \; (0.078)$ & $-0.028 \; (0.079)$ & $0.280 \; (0.123)$  & $-0.113 \; (0.125)$ & $-0.055 \; (0.119)$ \\
$\beta_{12}$  & $-0.117 \; (0.075)$ & $0.151 \; (0.080)$  & $0.156 \; (0.077)$  & $-1.511 \; (0.115)$ & $-0.685 \; (0.112)$ & $-0.718 \; (0.116)$ \\
$\beta_{21}$  & $-0.385 \; (0.044)$ & $3.246 \; (0.070)$  & $-3.284 \; (0.070)$ & $-0.048 \; (0.051)$ & $4.132 \; (0.087)$  & $-4.411 \; (0.085)$ \\
$\beta_{22}$  & $0.080 \; (0.045)$  & $-1.532 \; (0.072)$ & $1.500 \; (0.070)$  & $-0.036 \; (0.051)$ & $-1.968 \; (0.088)$ & $2.047 \; (0.089)$  \\
$\alpha_{11}$ & $-0.040 \; (0.035)$ & $-0.703 \; (0.058)$ & $0.594 \; (0.052)$  & $0.031 \; (0.037)$  & $-0.993 \; (0.075)$ & $0.942 \; (0.077)$  \\
$\alpha_{12}$ & $0.159 \; (0.036)$  & $1.878 \; (0.059)$  & $-1.637 \; (0.055)$ & $-0.006 \; (0.036)$ & $2.839 \; (0.080)$  & $-2.511 \; (0.080)$ \\
$\alpha_{21}$ & $-0.028 \; (0.039)$ & $0.666 \; (0.058)$  & $-0.702 \; (0.062)$ & $0.054 \; (0.042)$  & $1.031 \; (0.084)$  & $-0.950 \; (0.077)$ \\
$\alpha_{22}$ & $0.045 \; (0.037)$  & $-1.822 \; (0.057)$ & $1.933 \; (0.061)$  & $-0.074 \; (0.044)$ & $-2.590 \; (0.084)$ & $2.615 \; (0.081)$  \\

  \bottomrule
\end{tabular}}
\caption{ARB and RMSE (in brackets) for longitudinal and state-parameter estimates with a sample size $N = 200$ and length of longitudinal sequences $T = 10$.}\label{tab:s2}
\end{table}

\begin{table}[htbp]
 \center
 \smallskip 
 \resizebox{1.0\columnwidth}{!}{%
 \setlength\tabcolsep{15pt}
\begin{tabular}{lcccccccc}
\toprule
Correlation & \multicolumn{4}{c}{$\rho_{12} = 0.3$} & \multicolumn{4}{c}{$\rho_{12} = 0.8$}\\\cmidrule(r){2-5}\cmidrule(r){6-9}
Scenario & \multicolumn{2}{c}{($\mathcal{N} - \mathcal{N}$)} & \multicolumn{2}{c}{($\mathcal{T} - \mathcal{T}$)} & \multicolumn{2}{c}{($\mathcal{N} - \mathcal{N}$)} & \multicolumn{2}{c}{($\mathcal{T} - \mathcal{T}$)}\\\cmidrule(r){2-3}\cmidrule(r){4-5}\cmidrule(r){6-7}\cmidrule(r){8-9}
& AIC & BIC & AIC & BIC & AIC & BIC & AIC & BIC \\
\hline
\\
\# of hidden states \\
\qquad \qquad 2 & 20 & 67 & 79 & 92 & 33 & 68 & 92 & 96 \\
\qquad \qquad 3 & 34 & 22 & 19 & 8 & 36 & 25 & 6 & 4 \\
\qquad \qquad 4 & 46 & 11 & 2 & 0 & 31 & 7 & 2 & 0 \\
\bottomrule
\end{tabular}}
\caption{Absolute frequency distributions of the selected hidden states $M$ via AIC and BIC, with a sample size $N = 200$ and length of longitudinal sequences $T = 10$, over $B=100$ Monte Carlo replications.}\label{tab:s3}
\end{table}

\end{document}